\documentclass[preprint,nofootinbib]{revtex4}
\usepackage{graphicx}

\def\lsim{\lesssim}
\def\gsim{\gtrsim}

\begin{document}
\preprint{
 {\vbox{
 \hbox{MADPH--04--1378}
 \hbox{hep-ph/0405072}}}}

\title{The Littlest Higgs boson at a photon collider}

\author{Heather E. Logan}
\email{logan@physics.wisc.edu}
\affiliation{Department of Physics, University of Wisconsin, 1150 University
Avenue, Madison, Wisconsin 53706 USA}

\begin{abstract}
We calculate the corrections to the partial widths of the light Higgs 
boson in the Littlest Higgs model due to effects of the TeV-scale physics.  
We focus on the loop-induced Higgs coupling to photon pairs, which is
especially sensitive to the effects of new particles running in
the loop.  This coupling can be probed with high precision at a photon 
collider in the process $\gamma \gamma \to H \to b \bar b$ for a light 
Higgs boson with mass 115 GeV $\leq M_H \lsim 140$ GeV.  Using future 
LHC measurements of the parameters of the Littlest Higgs model, 
one can calculate a 
prediction for this process, which will serve as a test of the model 
and as a probe for a strongly-coupled UV completion at the 10 TeV scale.
We outline the prospects for measuring these parameters with sufficient
precision to match the expected experimental uncertainty on 
$\gamma\gamma \to H \to b \bar b$.
\end{abstract}

\maketitle
\section{Introduction}

Understanding the mechanism of electroweak symmetry breaking (EWSB) is the 
central goal of particle physics today.  A full understanding of EWSB 
will include a solution to the hierarchy or naturalness problem --
that is, why the weak scale is so much lower than the Planck scale.
Whatever is responsible for EWSB and its hierarchy, it must manifest 
experimentally at or below the TeV energy scale.  

Our first glimpse at the EWSB scale came from the electroweak precision 
data from the CERN LEP collider, which is sensitive to the Higgs boson 
mass in the Standard Model (SM) via radiative corrections.  This electroweak 
precision data points to the existence of a light Higgs boson in the 
SM, with mass below roughly 200 GeV \cite{LEPEWWG}.

The TeV scale is currently being probed at the Fermilab Tevatron 
and will soon be thoroughly explored at the CERN Large Hadron Collider 
(LHC).  Further into the future, a linear $e^+e^-$ collider will offer 
an excellent opportunity to study the dynamics of the new physics with 
uniquely high precision.  
The wealth of data on TeV scale physics promised by this experimental
program has driven model-building on the theoretical side.

A wide variety of models have been introduced over the past three 
decades to address EWSB and the hierarchy problem: supersymmetry, extra 
dimensions, strong dynamics leading to a composite Higgs boson, and 
the recent ``little Higgs'' models \cite{Littlest,LHModels}
in which the Higgs is a pseudo-Goldstone 
boson.  In this paper we consider the last possibility.  For concreteness,
we choose a particular model framework, the ``Littlest Higgs'' 
\cite{Littlest}, for our calculations.

In the little Higgs models, the SM Higgs doublet appears as a 
pseudo-Goldstone boson of an approximate global symmetry that is 
spontaneously broken at the TeV scale.  The models are constructed
as nonlinear sigma models, which become strongly coupled (and thus 
break down) no more than one loop factor
above the spontaneous symmetry breaking scale.  In fact, in many 
models unitarity 
violation in longitudinal gauge boson scattering appears to occur only
a factor of a few above the spontaneous symmetry breaking scale, due to
the large multiplicity of Goldstone bosons \cite{Hong-Jian}.  
Thus the little Higgs
models require an ultraviolet (UV) completion at roughly the 10 TeV 
scale.  The first UV completions of little Higgs models have been
constructed in Refs.~\cite{UVcompletion,turtles}.

The explicit breaking of the global symmetry, by gauge,
Yukawa and scalar interactions, gives the Higgs a mass and non-derivative
interactions, as required of the SM Higgs doublet.  The little Higgs models are
constructed in such a way that no \emph{single} interaction breaks 
\emph{all} of the symmetry forbidding a mass term for the SM Higgs doublet.  
This guarantees the cancellation of the one-loop quadratically divergent
radiative corrections to the Higgs boson mass.  Quadratic
sensitivity of the Higgs mass to the cutoff scale then arises only at the 
\emph{two}-loop level, so that a Higgs mass at the 100 GeV scale, two 
loop factors below the 10 TeV cutoff, is natural.

A light Higgs boson is the central feature of the little Higgs models.
In the Littlest Higgs model, the couplings of the Higgs boson to SM 
particles receive corrections due to the new TeV-scale particles
\cite{LHpheno,LHloop,Kilian,Deandrea}.
These corrections are suppressed by the square of the ratio of
the electroweak scale to the TeV scale, and are thus parametrically 
at the level of a few percent.  Percent-level measurements of Higgs
couplings are expected to be possible at a future linear $e^+e^-$ 
collider and its photon collider extension.

Corrections to the Higgs couplings can also be induced by the UV completion
at 10 TeV.  For example, the loop-induced Higgs coupling to photon pairs 
receives corrections from new heavy particles running in the loop.
If the UV completion is weakly coupled, these corrections should naively
be suppressed by the square of the ratio of the electroweak scale to 
the 10 TeV scale, and thus be too small to detect with the expected 
experimental capabilities.  However, if the UV completion is strongly
coupled, the strong-coupling enhancement counteracts the suppression from
the high mass scale, leading to corrections naively of the same order as
those from the TeV scale physics.
To reiterate, if the UV completion is weakly coupled, we expect the 
corrections to the Higgs couplings to be accurately predicted by the 
TeV-scale theory alone.  However, if the UV completion is strongly 
coupled, we expect the Higgs couplings to receive corrections from the
UV completion at the same level as the corrections from the TeV-scale
theory.

The parameters of the Littlest Higgs model can be measured at the LHC and
then used to calculate predictions for the corrections to the Higgs couplings
due to the TeV-scale physics.  Comparing these predictions to high-precision
Higgs coupling measurements will serve as a test of the model, as well as a
probe for a strongly-coupled UV completion.  In this paper, we focus on
the process $\gamma \gamma \to H \to b \bar b$, the rate for which will 
be measured with high precision at a future photon collider.

This paper is organized as follows.  We begin in Sec.~\ref{sec:expt}
with a brief review of the experimental prospects and a general discussion 
of the bounds that can be put on the dimension-six operator that generates 
a non-SM Higgs coupling to photon pairs.
In Sec.~\ref{sec:LHmodel} we outline the Littlest Higgs model \cite{Littlest}, 
following the notation of Refs.~\cite{LHpheno,LHloop}.
In Sec.~\ref{sec:corrections} we calculate the corrections to the Higgs 
couplings due to the TeV-scale new physics in the Littlest Higgs model,
focusing on the correction to $\gamma\gamma \to H \to b \bar b$.

In order to make predictions for the Higgs couplings, the TeV-scale
model parameters must be measured.  In Sec.~\ref{sec:LHC} we estimate 
the precision with which the parameters of the TeV-scale
theory must be measured at the LHC in order to give theoretical
predictions that match the precision of the photon collider measurement,
and discuss the prospects for doing so.
In Sec.~\ref{sec:uncert} we address the additional sources of 
experimental and theoretical uncertainty that affect our probe of 
the model.
Section~\ref{sec:conclusions} is reserved for our conclusions.
Formulas for the coupling correction factors are collected in an Appendix.

\section{Higgs production at a photon collider}
\label{sec:expt}

\subsection{Experimental considerations}

If the Higgs boson is sufficiently SM-like, its discovery is guaranteed
at the LHC \cite{LHCTDRs}.  Its mass will be measured with high precision
\cite{LHCTDRs}, and
in addition, LHC measurements of Higgs event rates in 
various signal channels allow for the extraction of certain combinations
of Higgs partial widths at the $10-30\%$ level \cite{LHCHiggsMeas}.
A future $e^+e^-$ linear collider will measure the production cross section
of a light Higgs boson in Higgsstrahlung or $WW$ fusion with percent-level
precision, as well as the important branching fractions with few-percent
precision \cite{LCs,TESLATDR}.  
A photon collider, which can be constructed from a linear
$e^+e^-$ or $e^-e^-$ collider through Compton backscattering of lasers
from the $e^{\pm}$ beams, can also measure rates for Higgs production
(in two-photon fusion) and decay into certain final states with few-percent 
level precision 
\cite{ggAsner,LeptonPhoton,Ohgaki,AGG,Jikia,Krawczyk1,Krawczyk2,Rosca}.

In this paper we focus on the Higgs coupling measurements that can be
made at a photon collider.  Experimental studies of the expected precisions 
with which the rates for $\gamma\gamma \to H \to X$ can be measured have 
been done for various photon collider designs (NLC, TESLA, JLC, and 
CLICHE\footnote{CLICHE, or the CLIC Higgs Experiment \cite{ggAsner},
is a low-energy $\gamma\gamma$ collider based on CLIC 1 \cite{CLIC1}, 
the demonstration project for the higher-energy two-beam accelerator 
CLIC \cite{CLIC}.}); 
their results are summarized in Table~\ref{tab:rates}.
\begin{table}
\begin{tabular}{llcccc}
\hline \hline
Study & & $M_H$ & $b \bar b$ & $W W^*$ & $\gamma\gamma$ \\
\hline
CLICHE & \cite{ggAsner,LeptonPhoton} \ & 115 GeV \ & 2\% & 5\% 
        & 22\% \\
JLC & \cite{Ohgaki} & 120 GeV \ & 7.6\% & -- & -- \\
NLC & \cite{AGG} & 120 GeV \ & 2.9\% & -- & -- \\
         &      & 160 GeV \ & 10\%  & -- & -- \\
TESLA & \cite{Jikia,Krawczyk1,Krawczyk2,Rosca} & 120 GeV \ & 1.7--2\% 
		& -- & -- \\
 & \cite{Krawczyk2} & 130 GeV \ & 1.8\% & -- & -- \\
 &                  & 140 GeV \ & 2.1\% & -- & -- \\
 &		    & 150 GeV \ & 3.0\% & -- & -- \\
 & \cite{Jikia,Krawczyk2}  & 160 GeV \ & 7.1--10\%  
		& -- & -- \\
\hline \hline
\end{tabular}
\caption{Expected experimental precision of the rate measurement of 
$\gamma\gamma \to H \to X$.  Dashes indicate that the corresponding 
study has not been done.  Not included are studies of Higgs boson decays 
to $WW$, $ZZ$ \cite{KrawczykHighM} and $t\bar t$ \cite{Asakawa} for
$M_H \geq 200$ GeV.
}
\label{tab:rates}
\end{table}
All the studies assume roughly one year's running at design luminosity.
The variations in results between different studies at the same
Higgs mass are believed
to be due mostly to the different photon beam spectra and luminosities 
at the different machines.
In all cases $\sqrt{s_{ee}}$ and the electron and laser 
polarizations have been optimized for maximum Higgs production.

From Table~\ref{tab:rates} we take away two lessons: (1) the rate for
$\gamma\gamma \to H \to b \bar b$ can be measured to about 2\% for a
SM-like Higgs boson with 115 GeV $\leq M_H \lsim 140$ GeV, and 
(2) this precision 
is better than will be obtained for any other Higgs decay mode for a Higgs 
boson in this mass range.

\subsection{Probing the $\gamma\gamma H$ coupling}

In the SM, the $\gamma\gamma H$ coupling arises from the loop-induced
dimension-6 operator
\begin{equation}
        \mathcal{L} = \frac{C}{\Lambda^2} h^{\dagger} h F^{\mu\nu} F_{\mu\nu},
        \label{eq:dim6op}
\end{equation}
where $h$ is the Higgs doublet, $F^{\mu\nu}$ is the electromagnetic
field strength tensor, $\Lambda$ is the mass scale that characterizes the
interaction, and $C$ is a dimensionless coefficient.  
This operator leads to the Higgs boson partial width
into photon pairs,
\begin{equation}
        \Gamma_{\gamma} = \frac{C^2 v^2 M_H^3}{2 \pi \Lambda^4},
\end{equation}
where $v = 246$ GeV is the SM Higgs vacuum expectation value (vev) 
and $M_H$ is the physical Higgs mass.

Taking, e.g., $M_H = 115$ GeV, we compute the partial width 
$\Gamma_{\gamma}$ using HDECAY \cite{HDECAY} to be
\begin{equation}
        \Gamma_{\gamma} = 6.65 \times 10^{-6} \ {\rm GeV}.
\end{equation}
This leads to the following estimate for the scale $\Lambda$ for the
SM loops that give rise to the $\gamma\gamma H$ coupling, for various
choices of 
$C$:\footnote{The dimension-6 coupling in Eq.~(\ref{eq:dim6op})
can only arise via loops, not through tree-level exchange of new heavy
particles, and by gauge invariance the photon always couples proportional 
to $e$.  Thus the value of $C$ corresponding to strongly-coupled
new physics is not of order $(4\pi)^2$ as would be estimated
using Naive Dimensional Analysis \cite{NDA} for strongly coupled 
tree-level exchange.  Instead, $C$ can be written in the form
$C \sim N^2 g_H^2 e^2 / 16 \pi^2 = N^2 \alpha_{EM} (\alpha_H/4\pi) (4\pi)$,
where $N$ counts the multiplicity of the particles in the loop,
$g_H$ is the Higgs coupling to the particles in the loop, $e^2$ accounts
for the photon couplings, and $1/16\pi^2$ is the loop factor.
For strong interactions, $\alpha_H/4\pi$ is of order one, so that $C$ is
of order $N^2 \alpha_{EM} (4\pi) \sim 0.1 N^2$.  Because the global 
symmetry groups in little Higgs models are typically rather large, 
their UV completions can be expected to have a large multiplicity
of charged particles at the UV cutoff (see, e.g., Ref.~\cite{UVcompletion}), 
leading to $C$ of order one for a strongly-coupled UV completion.
}
\begin{equation}
        \Lambda_{\rm SM} = \left\{ \begin{array}{rl}
                6.8 \ {\rm TeV} \ \ & C = 1 \\
                550 \ {\rm GeV} \ \ & C = 1/16\pi^2 \\
                170 \ {\rm GeV} \ \ & C = e^2/16\pi^2.
                \end{array} \right.
        \label{eq:LambdaSM}
\end{equation}
The SM coupling is generated primarily by $W$ boson and top quark loops,
with a characteristic energy scale around the weak scale.  This shows
the importance of the loop suppression and electromagnetic coupling
suppression of the operator in Eq.~(\ref{eq:dim6op}).

If new physics beyond the SM contributes to the $\gamma\gamma H$ 
coupling, we can parameterize its effect in Eq.~(\ref{eq:dim6op}) through
\begin{equation}
        \frac{C}{\Lambda^2} \to \frac{C_{\rm SM}}{\Lambda^2_{\rm SM}}
                + \frac{C_{\rm new}}{\Lambda^2_{\rm new}}.
\end{equation}
With the assumption that 
$C_{\rm SM}/\Lambda_{\rm SM}^2 \gg C_{\rm new}/\Lambda_{\rm new}^2$,
we can write the new physics correction relative to the SM partial width
as
\begin{equation}
        \frac{\delta \Gamma_{\gamma}}{\Gamma_{\gamma}}
        \simeq 2 \left| \frac{C_{\rm new}}{C_{\rm SM}} \right|
        \frac{\Lambda^2_{\rm SM}}{\Lambda^2_{\rm new}}.
\end{equation}

As in Eq.~(\ref{eq:LambdaSM}), the scale $\Lambda_{\rm new}$ that can
be probed with a measurement of $\Gamma_{\gamma}$ depends on the 
assumption for $C_{\rm new}$.  We consider two possibilities: weakly-coupled
loops, $C_{\rm new} = e^2/16\pi^2$, and strongly-coupled loops, 
$C_{\rm new} = 1$.
Assuming that $\Gamma_{\gamma}$ can be 
measured with 2\% precision, we find sensitivity to new physics 
scales at various confidence levels as given in Table~\ref{tab:Lambda-new}.
\begin{table}
\begin{tabular}{ccc}
\hline
\hline
   & \multicolumn{2}{c}{$\Lambda_{\rm new}$} \\
Confidence level \ & \ ($C_{\rm new} = e^2/16\pi^2$) \ 
        & \ ($C_{\rm new} = 1$) \ \\
\hline
$1\sigma$ & 1.7 TeV & 68 TeV \\
$2\sigma$  & 1.2 TeV & 48 TeV \\
$5\sigma$  & 0.74 TeV & 31 TeV \\
\hline
\hline
\end{tabular}
\caption{Sensitivity to the new physics scale $\Lambda_{\rm new}$ from a
2\% measurement of $\Gamma_{\gamma}$ at various confidence levels,
assuming the new physics is weakly coupled ($C_{\rm new} = e^2/16\pi^2$)
or strongly coupled ($C_{\rm new} = 1$).  $\Lambda_{\rm SM}$ was computed
for a 115 GeV SM Higgs boson using HDECAY \cite{HDECAY}.}
\label{tab:Lambda-new}
\end{table}
We find that the reach of this measurement for weakly-coupled new physics
is at the 1 TeV scale, while for strongly-coupled new physics it is at 
the few tens of TeV scale.

\section{The Littlest Higgs model}
\label{sec:LHmodel}

In this section we outline the Littlest Higgs model \cite{Littlest}
and define the parameters relevant for our analysis, following the notation
of Refs.~\cite{LHpheno,LHloop}.

The Littlest Higgs model consists of a nonlinear sigma model with a global
SU(5) symmetry which is broken down to SO(5) by a vacuum condensate
$f \sim$ TeV.  A subgroup 
SU(2)$_1 \times$SU(2)$_2 \times$U(1)$_1 \times$U(1)$_2$ of the global SU(5)
is gauged, with gauge couplings $g_1$, $g_2$, $g_1^{\prime}$ and 
$g_2^{\prime}$, respectively.  The breaking of the global SU(5) down to 
SO(5) by the condensate $f$ simultaneously breaks the gauge group down
to its diagonal SU(2)$\times$U(1) subgroup, which is identified as the
SM electroweak gauge group.  The breaking of the global symmetry gives
rise to 14 Goldstone bosons, four of which are eaten by the broken gauge
generators, leading to four vector bosons with masses of order $f$: an
SU(2) triplet, $Z_H$ and $W_H^{\pm}$, and a U(1) boson $A_H$.

Besides the condensate $f$, the heavy gauge boson sector is parameterized
in terms of two mixing angles,
\begin{equation}
        0 < c \equiv \cos\theta  = \frac{g_1}{\sqrt{g_1^2+g_2^2}} < 1, 
        \qquad \qquad
        0 < c^{\prime} \equiv \cos\theta^{\prime} 
        = \frac{g_1^{\prime}}{\sqrt{g_1^{\prime 2} + g_2^{\prime 2}}} < 1.
\end{equation}
We also define $s \equiv \sqrt{1-c^2}$ and 
$s^{\prime} \equiv \sqrt{1 - c^{\prime 2}}$.
The TeV-scale gauge boson masses are given to leading order in
$v^2/f^2$ in terms of these parameters by
\begin{equation}
  M_{Z_H} = M_{W_H} = \frac{gf}{2sc},
  \qquad \qquad
  M_{A_H} = \frac{g^{\prime}f}{2\sqrt{5}s^{\prime}c^{\prime}}.
  \label{eq:heavygaugemasses}
\end{equation}
The parameters $c$ and $c^{\prime}$ also control the couplings of
the heavy gauge bosons to fermions.\footnote{The couplings of $A_H$
to fermions are quite model-dependent, depending on the choice of the
fermion U(1) charges under the two U(1) groups \cite{LHpheno,GrahamEW2}.
For the corrections to the Higgs couplings, however, there is no model
dependence related to the choice of the $A_H$ couplings to fermions,
since $A_H$ only enters via its mixing with the $Z$ boson.  This mixing 
depends only on the Higgs doublet U(1) charges and is fixed by the 
model \cite{Littlest}.}

An alternate version of the model, which we will also consider, starts
with only SU(2)$_1 \times$SU(2)$_2 \times$U(1)$_Y$ gauged; this model contains
no $A_H$ boson.  Since the $A_H$ boson tends to cause significant 
custodial isospin
breaking and corrections to four-fermion neutral current interactions, 
this alternate version of the model is preferred by the electroweak precision 
data \cite{GrahamEW1,JoAnneEW,GrahamEW2}.  Since the $A_H$ is typically
also quite light, this version is also preferred by the direct exclusion
bounds from the Tevatron \cite{JoAnneEW,Park}.

The ten remaining uneaten Goldstone bosons transform under the SM gauge
group as a doublet $h$ (identified as the SM Higgs doublet) and a 
triplet $\phi$.\footnote{If only one U(1) is gauged so that the model 
contains no $A_H$ particle, then the spectrum contains an additional uneaten 
Goldstone boson that is an electroweak singlet pseudoscalar.  
We assume that this extra singlet does not mix with the SM-like Higgs 
boson $H$.}
The components $\Phi^{++}$, $\Phi^+$, $\Phi^0$ (scalar) and 
$\Phi^P$ (neutral pseudoscalar) of the triplet get a mass, to leading
order in $v^2/f^2$, of
\begin{equation}
  M_{\Phi} = \frac{\sqrt{2} M_H f}{v \sqrt{1-x^2}},
  \label{eq:MPhi}
\end{equation}
where $x$ is a free 
parameter of the Higgs sector proportional to the triplet vev $v^{\prime}$
and defined as
\begin{equation}
 0 \le x = \frac{4fv^\prime}{v^2} < 1.
\end{equation} 
The constraint $x < 1$ is required to obtain the correct electroweak symmetry
breaking vacuum and avoid giving a TeV-scale vev to the scalar triplet
(see Ref.~\cite{LHpheno} for further details).

Finally, the top quark sector is modified by the addition of a heavy
top-like quark $T$.
The top sector is parameterized by
\begin{equation}
        0 < c_t = \frac{\lambda_1}{\sqrt{\lambda_1^2 + \lambda_2^2}} < 1,
\end{equation}
where the dimensionless couplings $\lambda_{1,2}$ are defined according
to the normalization given in Ref.~\cite{LHpheno}.
Together with $f$, this parameter controls the $T$ mass (we also define 
$s_t \equiv \sqrt{1 - c_t^2}$),
\begin{equation}
  M_T = \frac{m_t f}{v s_t c_t}.
  \label{eq:MT}
\end{equation}
The parameter $c_t$ also controls the mixing between $t$ and $T$ at order
$v^2/f^2$, which generates a $TbW$ coupling leading to single $T$ 
production through $bW$ fusion at hadron colliders \cite{LHpheno,Peskin}.

\section{Corrections to Higgs observables}
\label{sec:corrections}

In any theory beyond the SM, corrections to SM observables must be 
calculated relative to the SM predictions for a given set of SM
electroweak inputs.  These electroweak inputs are usually taken to 
be the Fermi constant $G_F$ defined in muon decay, the $Z$ mass $M_Z$, 
and the electromagnetic fine structure constant $\alpha$.
Thus, a calculation of corrections to, e.g., Higgs couplings due to
new physics must necessarily involve a calculation of the corrections
to the SM electroweak input parameters due to the same new physics.

In the Littlest Higgs model, it is most straightforward to calculate 
the corrections to the Higgs couplings in terms of the SM Higgs vev
$v = 246$ GeV.  To obtain useful predictions of the couplings, this
must be related to the Fermi constant in the Littlest Higgs model
according to $v^{-2} = \sqrt{2} G_F y^2_{G_F}$, where 
$y^2_{G_F} = 1 + \mathcal{O}(v^2/f^2)$ 
is a correction factor given in the Appendix.

\subsection{Higgs partial widths}

In this section we present the formulas for the corrections to the
Higgs partial widths to SM particles.  We write the partial widths
$\Gamma_i$ in the Littlest Higgs model normalized to the corresponding
SM partial width, $\Gamma_i^{\rm SM}$.  The partial widths are 
written in terms of correction factors $y_i$, which are collected
in the Appendix.
For the SM electroweak inputs we take the parameters $G_F$, $M_Z$ 
and $\alpha$.

The corrections to the loop-induced partial widths of the Higgs boson 
into photon pairs and gluon pairs were computed in the Littlest Higgs 
model in Ref.~\cite{LHloop}; we list them in the Appendix for completeness.

The corrections to the tree-level couplings of the Higgs boson in 
the Littlest Higgs model can be derived to order $v^2/f^2$ from 
the couplings given in Ref.~\cite{LHpheno}.
The partial widths of the Higgs boson into $Z$ boson pairs ($\Gamma_Z$),
top quark pairs ($\Gamma_t$)\footnote{The Higgs coupling to the top quark 
gets a different correction than the Higgs couplings of the light fermions 
due to the mixing between $t$ and $T$ in the Littlest Higgs model.  The 
correction to $\Gamma_t$ is only important in Higgs decay if 
$M_H \gsim 2 m_t$.}, and pairs of other fermions
($\Gamma_f$) normalized to their SM values are given by
\begin{equation}
        \Gamma_Z / \Gamma_Z^{\rm SM} = y^2_{G_F} y^2_Z,
	\qquad \qquad
	\Gamma_t / \Gamma_t^{\rm SM} = y^2_{G_F} y^2_t,
        \qquad \qquad
        \Gamma_f / \Gamma_f^{\rm SM} = y^2_{G_F} y^2_f.
\label{eq:treeGammas}
\end{equation}

The correction to the partial width for the Higgs decay to $W$ bosons 
is a little subtle when $G_F$, $M_Z$ and $\alpha$ are
used as inputs because the relation between these inputs and the physical
$W$ boson mass receives corrections from the Littlest Higgs model.  
The partial width of $H \to WW^{(*)}$ depends on the $W$ mass in the 
kinematics, especially in the intermediate Higgs mass range, 
$115 \ {\rm GeV} \lsim M_H \lsim 2 M_W$.
To deal with this, we follow the same approach taken by the program HDECAY 
\cite{HDECAY} for the Minimal Supersymmetric Standard Model (MSSM), 
which is to define the $H \to WW^{(*)}$ partial width in the MSSM in terms
of the SM partial width simply by scaling by the ratio of the $WWH$ 
couplings-squared in the two models, 
ignoring the shift in the kinematic $W$ mass.  Thus, we calculate only
the correction to the coupling-squared in the Littlest Higgs model, and do not 
worry about the shift due to the $W$ mass correction in the kinematics.
We find,
\begin{equation}
        \Gamma_W / \Gamma_W^{\rm SM} = y^2_{G_F} y^2_W 
        \frac{y^4_{M_W}}{y^4_{M_Z}} y^4_{c_W}.
        \label{eq:GW/SM}
\end{equation}

The corrections to the Higgs couplings involved in 
$\Gamma_Z/\Gamma_Z^{\rm SM}$, $\Gamma_f/\Gamma_f^{\rm SM}$, 
$\Gamma_t/\Gamma_t^{\rm SM}$, and $\Gamma_W/\Gamma_W^{\rm SM}$ in the 
Littlest Higgs model were derived previously in Ref.~\cite{Kilian}
by integrating out the heavy degrees of freedom; we agree with their
results.

\subsection{Higgs production and decay}

The partial width ratios given above can immediately be used to find
the corrections to the Higgs boson production cross sections in gluon
fusion and in two-photon fusion, since the production cross section
is simply proportional to the corresponding Higgs partial width.  
Detailed results were given in Ref.~\cite{LHloop}.  
For other Higgs boson production channels, 
the cross section corrections are more complicated because in addition
to the corrections to the Higgs couplings to SM particles, exchange
of the TeV-scale particles in the production diagrams must also be
taken into account \cite{LHHiggsprod}.  This is beyond the scope of
our current work; we thus focus on Higgs production in two-photon 
collisions.\footnote{We do not consider Higgs production in gluon fusion 
here because the large SM theoretical uncertainty from QCD corrections
is likely to hide the corrections due to new TeV-scale physics.
The QCD corrections to Higgs production in gluon fusion have been computed
at next-to-next-to-leading order \cite{NNLOggH}.  The remaining
renormalization and factorization scale uncertainty due to uncomputed
higher-order QCD corrections is at the 15\% level.}

The Higgs decay branching ratio to a final state $X$, 
${\rm BR}(H \to X) = \Gamma_X / \Gamma_{\rm tot}$, is computed in
terms of the SM branching ratio as follows:
\begin{equation}
        \frac{{\rm BR}(H \to X)}{{\rm BR}(H \to X)^{\rm SM}}
        = \frac{\Gamma_X / \Gamma_X^{\rm SM}}
        {\Gamma_{\rm tot} / \Gamma_{\rm tot}^{\rm SM}}.
\end{equation}
The numerator can be read off from Eqs.~(\ref{eq:Gga/SM}--\ref{eq:GW/SM}).
The denominator requires a calculation of the Higgs total width, which
we perform as follows.  We compute the Higgs partial width into each 
final state for a given Higgs mass in the SM using HDECAY \cite{HDECAY}.  
The SM total width $\Gamma_{\rm tot}^{\rm SM}$ is of course the sum of
these partial widths.  We then find the total width in the Littlest Higgs
model by scaling each partial width in the sum by the appropriate ratio
from Eqs.~(\ref{eq:treeGammas}--\ref{eq:GW/SM}) and 
(\ref{eq:Gga/SM}--\ref{eq:Gg/SM}).

A quick examination of the corrections to the Higgs partial widths
given above reveals that the corrections to the $\gamma\gamma \to H$ 
production cross section and to all of the Higgs branching
ratios are parametrically of order $v^2/f^2$.  In particular, no 
coupling receives especially large corrections.  This is in contrast
to the MSSM, in which the corrections to the couplings of the light 
SM-like Higgs boson to down-type fermions are parametrically larger
than those to up-type fermions or to $W$ and $Z$ bosons \cite{MSSMHiggs}.
Thus in the Littlest Higgs model there is no ``golden channel'' in which 
we expect to see especially large deviations from the SM Higgs couplings.
We therefore expect the experimentally best-measured channel to give 
the highest sensitivity to TeV-scale effects.  For that reason, in the
rest of this paper we focus on the channel 
$\gamma\gamma \to H \to b \bar b$.  We take the Higgs mass $M_H = 115$ GeV
in our numerical calculations.  Changing the Higgs mass has only a small
effect on the size of the corrections to the Higgs couplings; however,
it affects the precision with which the rate for 
$\gamma\gamma \to H \to b \bar b$ can be measured.

\begin{figure}
\resizebox{0.5\textwidth}{!}{
\rotatebox{270}{\includegraphics[50,50][555,600]{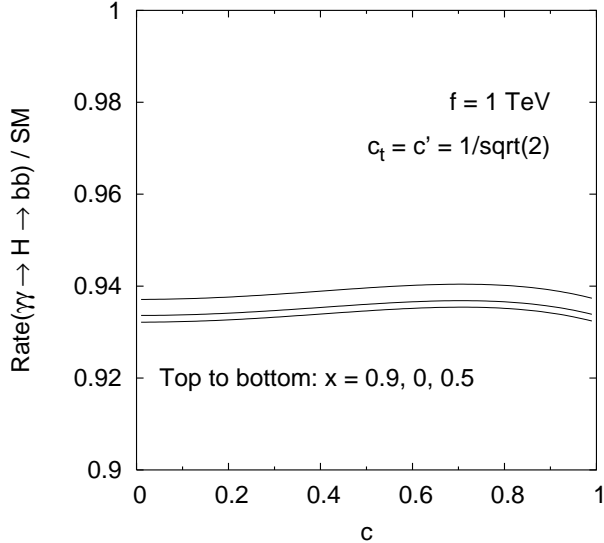}}}
\caption{Rate for $\gamma\gamma \to H \to b \bar b$, normalized to its 
SM value, as a function of $c$ for $x = 0$, 0.5 and 0.9 (solid
lines).
The other parameters are $f = 1$ TeV, 
$c_t = c^{\prime} = 1/\sqrt{2}$, and $M_H = 115$ GeV.
}
\label{fig:rate-curves}
\end{figure}

In Fig.~\ref{fig:rate-curves} we plot the rate for 
$\gamma\gamma \to H \to b \bar b$, normalized to its SM value, 
as a function of $c$ for various values of $x$, with $f = 1$ TeV and 
$c_t = c^{\prime} = 1/\sqrt{2}$.
As far as the Higgs couplings are concerned,
the choice $c^{\prime} = 1/\sqrt{2}$ is equivalent to removing the
$A_H$ boson from the model.  
Defining the rate for $\gamma\gamma \to H \to b \bar b$ in the Littlest
Higgs model as $R = R_{\rm SM} + R_{\rm LH}$, the deviation 
$R_{\rm LH}/R_{\rm SM}$
of the rate from its SM value scales with $f$ as $1/f^2$, for fixed
values of $c$, $c^{\prime}$, $x$ and $c_t$.

We see that the correction due to the TeV-scale new physics is roughly
$-6\%$ for $f= 1$ TeV, and depends only weakly on the parameters
$c$ and $x$.\footnote{The remaining parameter dependence
will be discussed in the next section.}  A 2\% measurement of the
rate for $\gamma\gamma \to H \to b \bar b$ thus gives a non-trivial
test of the model.

\section{Measuring the input parameters}
\label{sec:LHC}

In order to predict the corrections to the Higgs couplings due to
the TeV-scale physics in the Littlest Higgs model, one must measure
the five independent free parameters of the model.
There are two natural choices for the set of input parameters:
\begin{enumerate}
\item $c_t$, $x$, $f$, $c$, $c^{\prime}$, and
\item $c_t$, $x$, $f$, $M_{Z_H}$, $M_{A_H}$.
\end{enumerate}
The correction factors $y_i$ in the Appendix have been given 
in both parameterizations.  From the formulas in the Appendix it 
is easy to see that in the first parameterization, the dependence
on each of the variables $c$, $c^{\prime}$, $c_t$ and $x$ is independent,
while the $f$ dependence is an overall $1/f^2$ scaling.
In the second parameterization, the dependence on $M_{Z_H}$ and $M_{A_H}$
separates from the other variables (including $f$); the dependence
on $c_t$ is independent from that on $x$, with both scaled by $1/f^2$
as before.

The sensitivity of our test of the Littlest Higgs model and the reach 
of our probe of its UV completion are limited by the experimental 
uncertainty in the photon collider measurement of 
$\gamma\gamma \to H \to b \bar b$.  Ideally, we would like the 
theoretical uncertainty in our prediction for 
$\gamma\gamma \to H \to b \bar b$ in the Littlest Higgs model to
be smaller than this experimental uncertainty.  This theoretical 
uncertainty comes from 
uncertainties in the input parameters, which we assume will be 
measured at the LHC through the properties of the TeV-scale 
particles.\footnote{We address additional sources of uncertainty in
Sec.~\ref{sec:uncert}.}
We therefore 
study the sensitivity of the prediction for $\gamma\gamma \to H \to b \bar b$
to each of the input parameters.  This allows us to estimate whether
the LHC measurements will allow a prediction for 
$\gamma\gamma \to H \to b \bar b$ with precision comparable to that of 
the photon collider measurement.

We choose as our standard of precision a 1\% uncertainty in 
$\delta R/R_{\rm SM}$.  Four such parametric uncertainties added 
in quadrature match the expected 2\% experimental uncertainty.
The desired precision $\delta X/X$ on parameter $X$ scales linearly
with the precision on $\delta R/R_{\rm SM}$, so the results shown below
can be scaled for other precision requirements.  We take $M_H = 115$ GeV
in our numerical calculations.  Because $R_{\rm LH}/R_{\rm SM} \sim -6\%$
for $f=1$ TeV, $R_{\rm LH}$ need only be calculated to 15\% precision 
to obtain an overall 1\% uncertainty on $R$.  Parameter measurements at
this level of precision are feasible at the LHC.

We first consider the two dimensionless parameters, $c_t$ and $x$.  
The dependence of the Higgs partial widths on these two parameters is
the same in either of the parameterizations given above.  The
precision with which $c_t$ and $x$ must be measured for a given
$\delta R/R_{\rm SM}$ depends on the scale parameter $f$.  

\begin{figure}
\resizebox{0.5\textwidth}{!}{\rotatebox{270}{
\includegraphics[50,50][555,600]{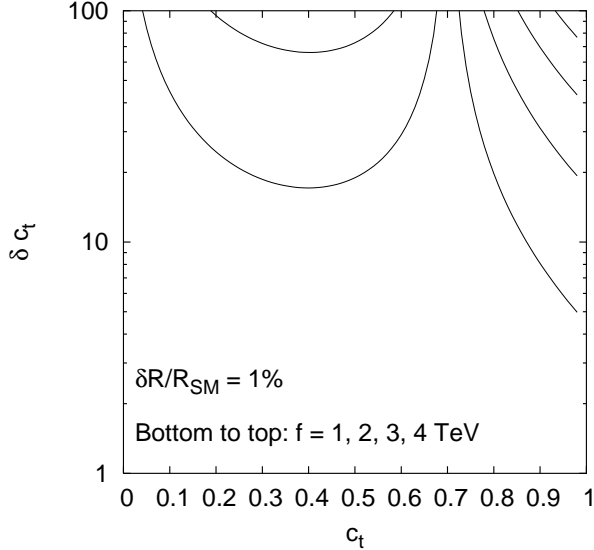}}}
\caption{Precision on $c_t$ required for $\delta R/R_{\rm SM} = 1\%$.
The solid lines are for $f = 1$, 2, 3, 4 TeV (bottom to top).}
\label{fig:ct}
\end{figure}

In Fig.~\ref{fig:ct}
we show the precision with which $c_t$ must be measured to give
$\delta R/R_{\rm SM} = 1\%$.
Even for low $f = 1$ TeV, the precision $\delta c_t$ required to give
$\delta R/R_{\rm SM} = 1\%$ is greater than one, meaning that no 
measurement of this parameter is required.  It is easy to understand
why the $c_t$ dependence of $\gamma\gamma \to H \to b \bar b$ is so 
weak.  A quick examination of the $y_i$ factors in the Appendix shows
that the $c_t$ dependence enters only through the Higgs couplings to
$t$ and $T$.  For the Higgs mass of 115 GeV that we consider, 
$H \to t \bar t$ is kinematically forbidden, so that the $c_t$ parameter
enters only through the $t$ and $T$ loops in $\Gamma_{\gamma}$ (which
controls the production cross section and affects the Higgs total width
at the per-mil level) and to a small extent $\Gamma_g$
(which enters the Higgs total width).  The $c_t$ dependence 
of $\Gamma_{\gamma,g}$ is very weak \cite{LHloop} because it
enters proportional to the 
difference between $F_{1/2}(\tau_t)$ and $F_{1/2}(\tau_T)$ in 
Eqs.~(\ref{eq:Gga}-\ref{eq:Gg/SM}):
\begin{equation}
	\sum_{i=t,T} y_i F_{1/2}(\tau_i) = \cdots +
	\frac{v^2}{f^2} c_t^2 s_t^2
	\left[ F_{1/2}(\tau_t) - F_{1/2}(\tau_T) \right].
	\label{eq:F12diff}
\end{equation}
In the limit $m_i \gg M_H$, $F_{1/2}(\tau_i) \to -4/3$.
For $M_H = 115$ GeV, this heavy-quark limit is already a good 
approximation for the top quark; in particular, for $m_t = 175$ GeV,
$F_{1/2}(\tau_t)$ differs from the heavy-quark limit by only 2.6\%,
leading to a large cancellation in Eq.~(\ref{eq:F12diff}).
For larger $M_H$ values, the $c_t$ dependence will become more important;
however, even for $M_H \sim 200$ GeV, $F_{1/2}(\tau_t)$ differs from the 
heavy-quark limit by less than 10\%.

If a measurement of $c_t$ were desired, it could be obtained from
the $T$ production cross section in $Wb$ fusion \cite{LHpheno,Peskin}
or from the $T$ mass as given in Eq.~(\ref{eq:MT}).

\begin{figure}
\resizebox{\textwidth}{!}{\rotatebox{270}{
\includegraphics[50,50][555,600]{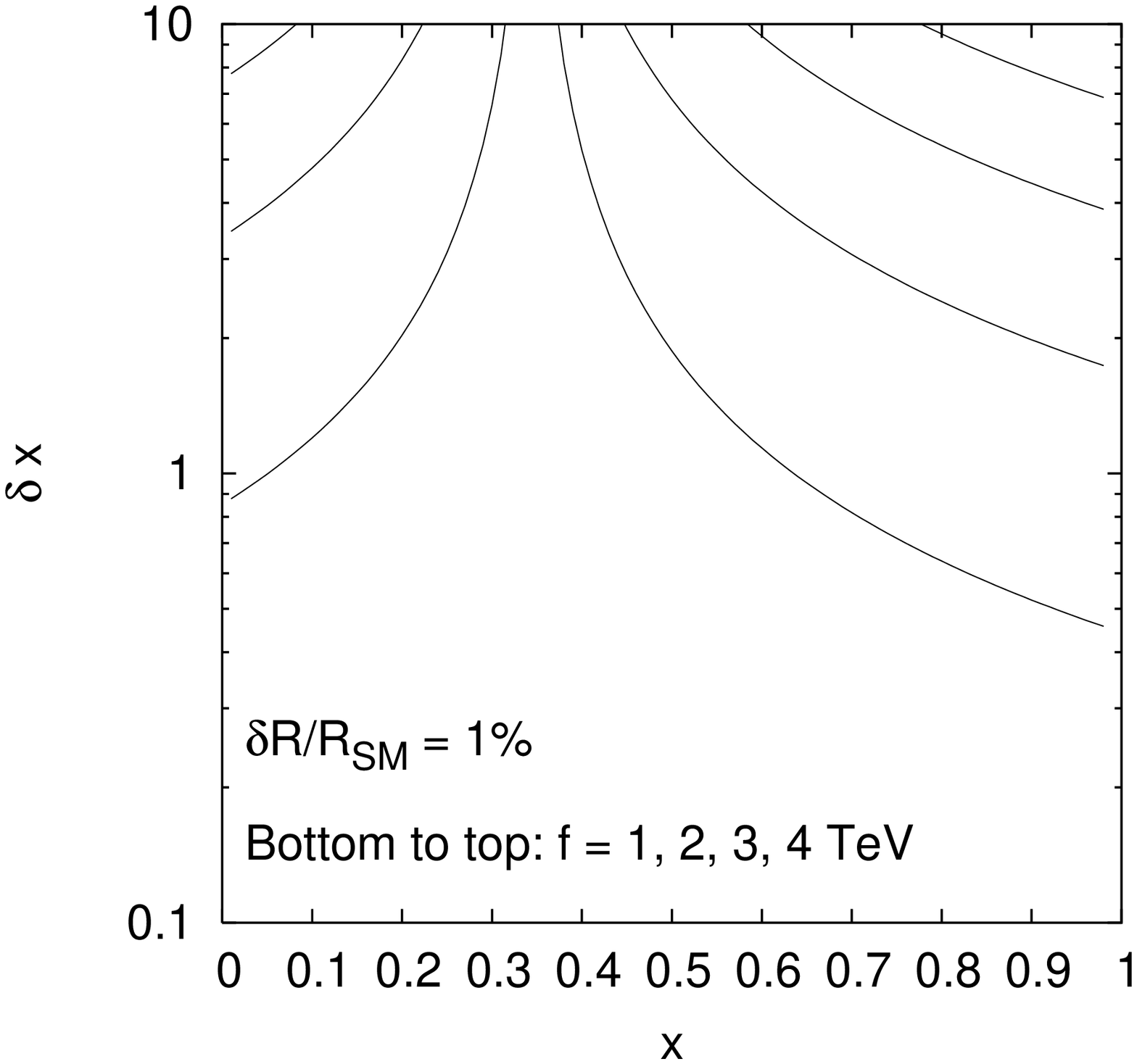}}
\rotatebox{270}{
\includegraphics[50,50][555,600]{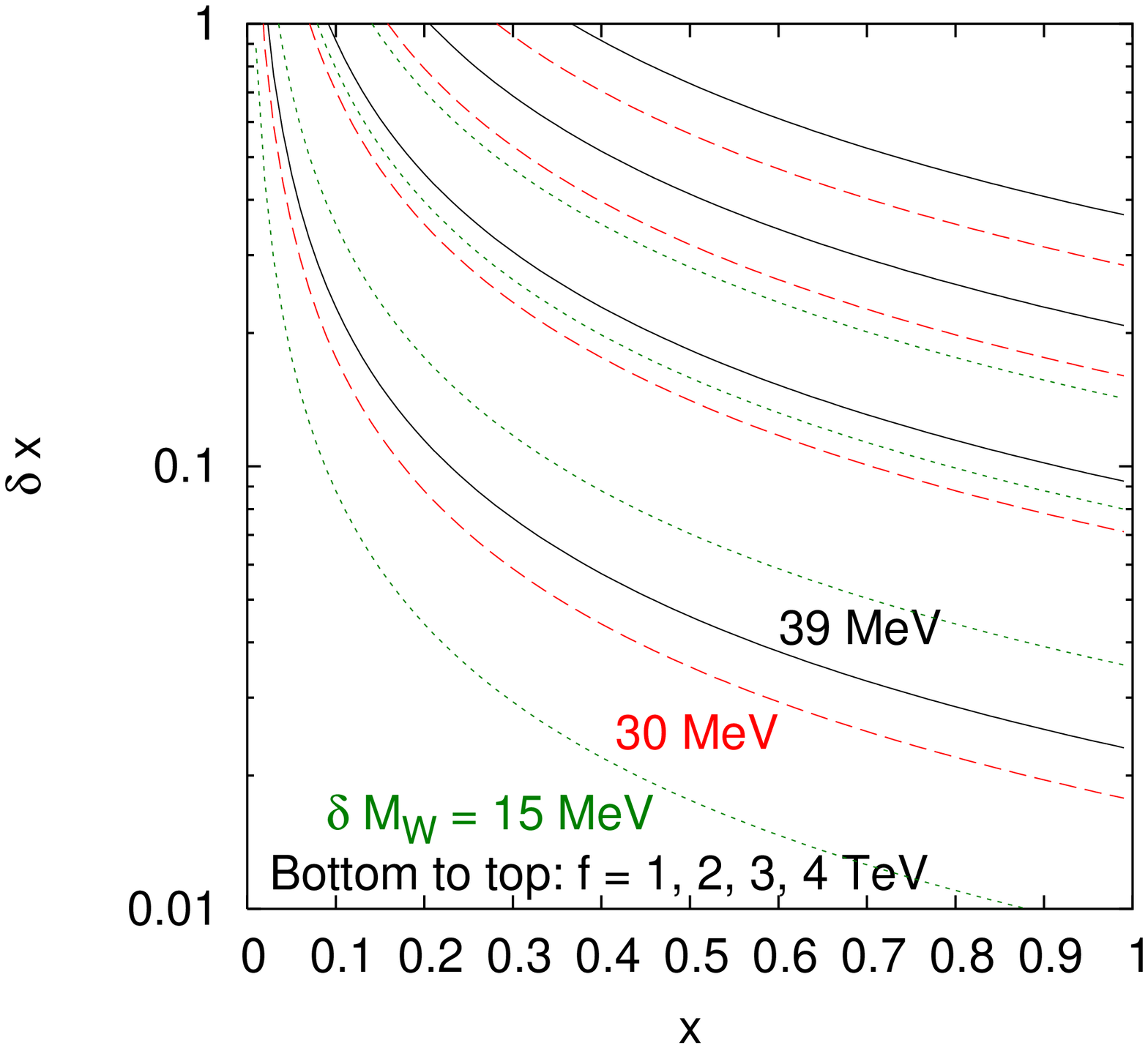}}}
\caption{Left: Precision on $x$ required for $\delta R/R_{\rm SM} = 1\%$.
The solid lines are for $f = 1$, 2, 3, 4 TeV (bottom to top).
Right: Precision on $x$ obtainable from $M_W$ measurement once
the other model parameters are known, for the current precision,
$\delta M_W = 39$ MeV (black solid lines), and the goals for Tevatron Run II
(2 fb$^{-1}$), $\delta M_W = 30$ MeV (red long-dashed lines),
and the LHC (10 fb$^{-1}$), $\delta M_W = 15$ MeV (green short-dashed lines),
with $f = 1$, 2, 3, 4 TeV (bottom to top for each line type).
Note the different scales on the $y$-axis.}
\label{fig:x}
\end{figure}

In the left panel of Fig.~\ref{fig:x} we show the precision with which
$x$ must be measured to give $\delta R/R_{\rm SM} = 1\%$.
A rough measurement of this parameter is needed if $f$
is relatively low.

The ideal place to measure $x$ is in the scalar triplet sector.  The 
mass of the scalar triplet depends on $x$ as given in Eq.~(\ref{eq:MPhi}).
The doubly-charged member of the scalar triplet, $\Phi^{++}$, can also
be produced in resonant like-sign $WW$ scattering, 
$W^+W^+ \to \Phi^{++} \to W^+W^+$ \cite{LHpheno} with a cross section
proportional to $x^2 v^4/f^2$.  Unfortunately, the cross section is
quite small because of the $v^2/f^2$ suppression, and is not likely to
be visible above background \cite{ATLASLH}.

Alternatively, $x$ can be measured through its effects on electroweak
precision observables \cite{Kilian}.  
We consider for example the $W$ boson mass.
The $W$ boson mass receives a correction in the Littlest Higgs model
given at tree level to order $v^2/f^2$ by
\begin{eqnarray}
        M_W^{\rm LH} &=& 
	M_W^{\rm SM} \frac{y_{M_W} y_{c_W}}{y_{M_Z}} 
	\nonumber \\
	&=& M_W^{\rm SM}
        \left\{ 1 + \frac{v^2}{2 f^2} \left[
        - \frac{s_W^2}{c_W^2 - s_W^2} c^2 s^2
        + \frac{c_W^2}{c_W^2 - s_W^2} \left( 
                \frac{5}{4} \left( c^{\prime 2} - s^{\prime 2} \right)^2
                - \frac{1}{4} x^2 \right) \right] \right\}.
\end{eqnarray}
If the parameters $f$, $c$, and $c^{\prime}$ (alternatively $f$, $M_{Z_H}$,
and $M_{A_H}$) are known, $x$ can
be extracted from the measurement of $M_W$ with a precision given by
\begin{equation}
        \delta x = \frac{\delta M_W}{M_W^{\rm SM}} \frac{4 f^2}{v^2 x}
        \frac{c^2_W - s^2_W}{c^2_W}.
\end{equation}
This precision is shown in the right panel of Fig.~\ref{fig:x} for the
current $M_W$ measurement, $\delta M_W = 39$ MeV \cite{PDG}, and
for the expected precisions obtainable with 2 fb$^{-1}$ of data in 
Run II of the Tevatron, $\delta M_W = 30$ MeV (per experiment) 
\cite{MWBaur,MWTev}, and with 10 fb$^{-1}$ of data at the LHC,
$\delta M_W \simeq 15$ MeV (combining two experiments and multiple 
channels) \cite{MWBaur,MWLHC}.
Even the current $M_W$ measurement gives enough precision on $x$
to meet the requirement of $\delta R/R_{\rm SM} = 1\%$ if the 
parameters $f$, $c$, and $c^{\prime}$ are known, except for 
low $x \lsim 0.05$ for $f = 1$ TeV.

We next consider the scale parameter $f$.
The sensitivity of $\gamma\gamma \to H \to b \bar b$ to $f$ depends on 
the parameterization and the values of the other parameters.
In the first parameterization ($c_t$, $x$, $f$, $c$, $c^{\prime}$),
the sensitivity to $f$ depends on the parameters $x$, $c$ and $c^{\prime}$,
while in the second parameterization ($c_t$, $x$, $f$, $M_{Z_H}$, $M_{A_H}$),
the sensitivity to $f$ depends only on the parameter $x$.\footnote{We
ignore the parameter $c_t$ because the rate for 
$\gamma\gamma \to H \to b \bar b$ depends upon it only very
weakly, as shown in Fig.~\ref{fig:ct}.}
This is due to the parameter dependence of the terms multiplying 
$1/f^2$ in the expressions for $y_i$ given in the Appendix.

In Fig.~\ref{fig:f} we show the precision with which $f$ must be measured 
to give $\delta R/R_{\rm SM} = 1\%$ in the second parameterization.
\begin{figure}
\resizebox{\textwidth}{!}{\rotatebox{270}{
\includegraphics[50,50][555,600]{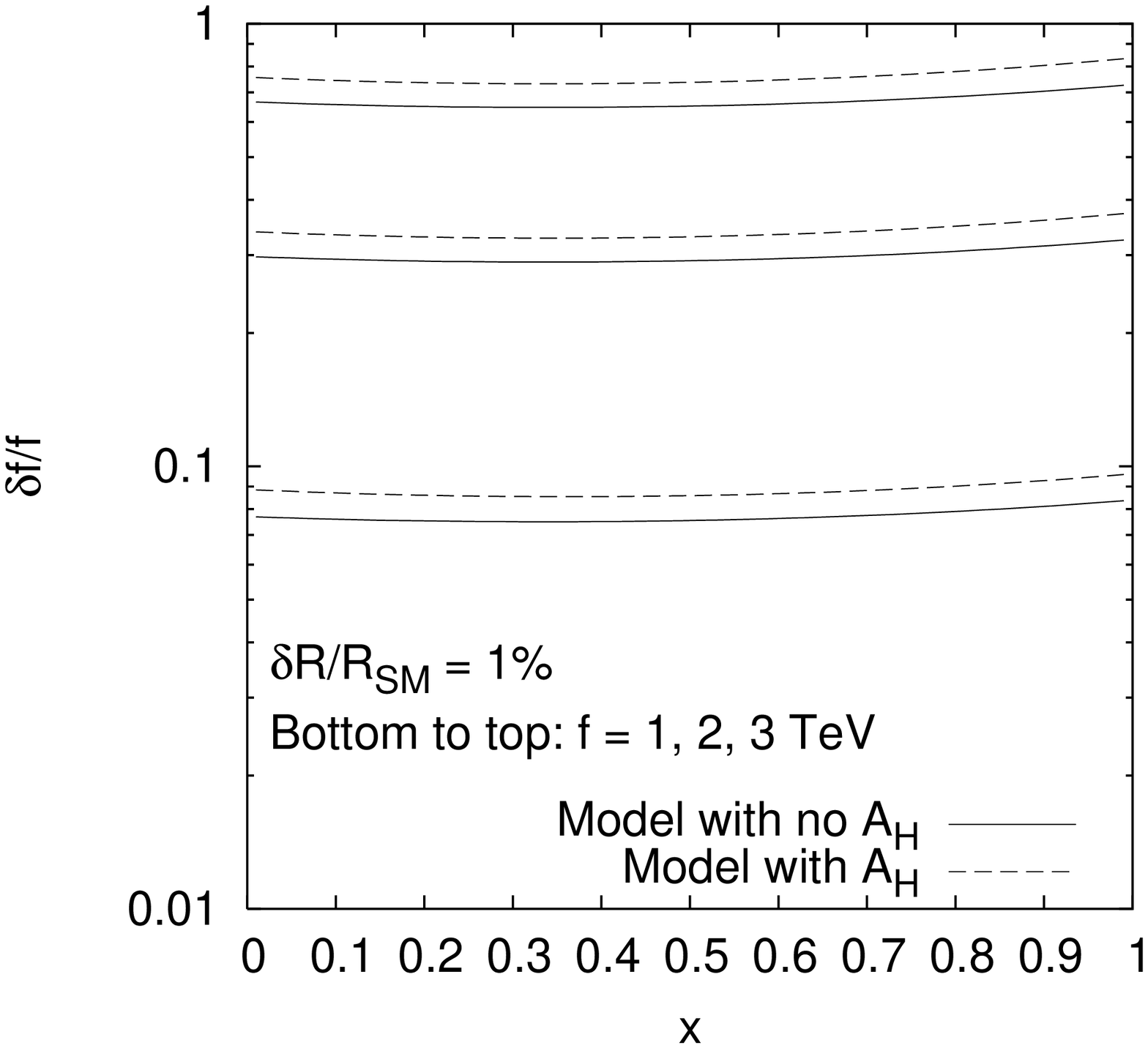}}
\rotatebox{270}{
\includegraphics[50,50][555,600]{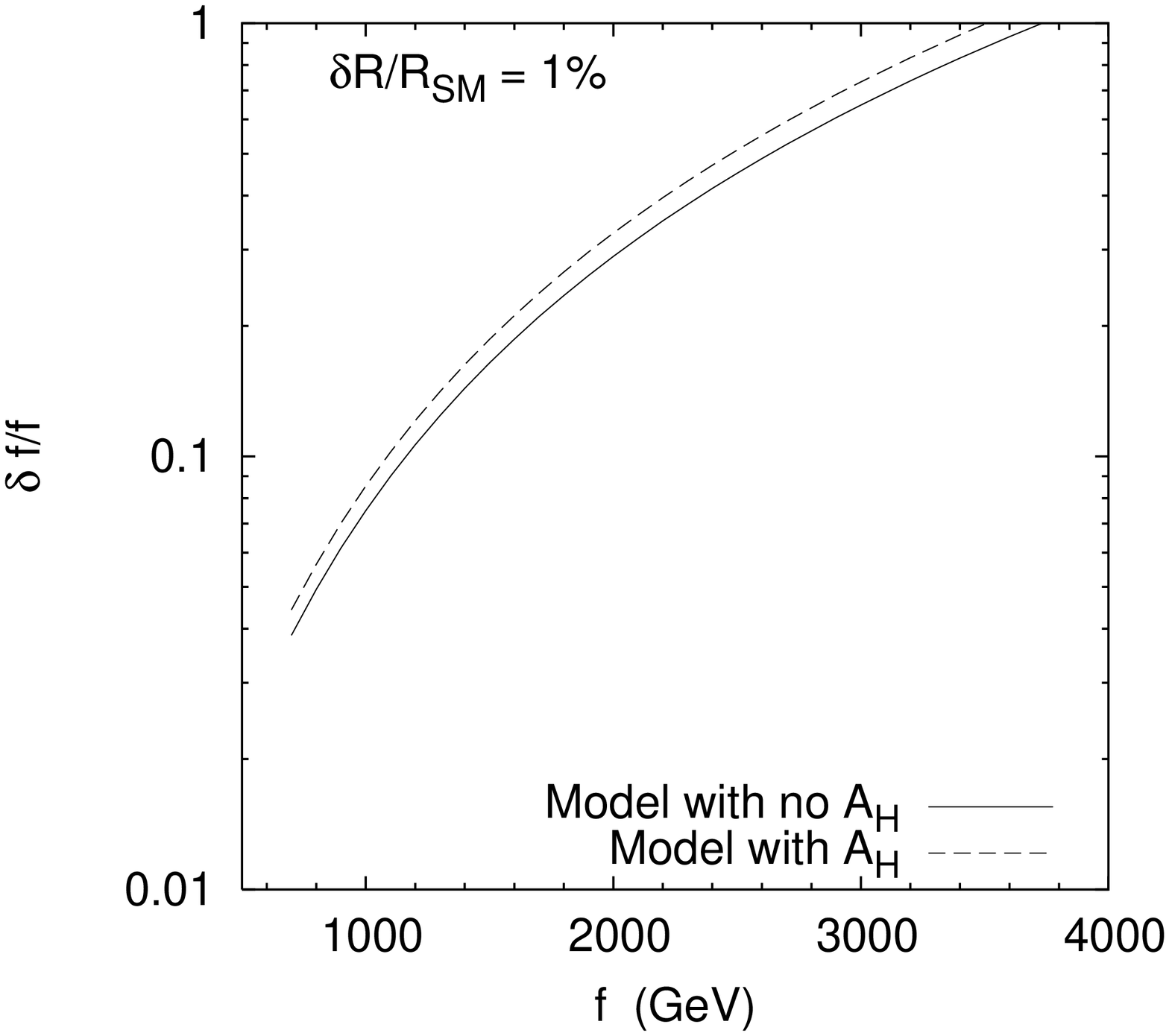}}}
\caption{Precision on $f$ required for $\delta R/R_{\rm SM} = 1\%$
in the second parameter set ($c_t$, $x$, $f$, $M_{Z_H}$, $M_{A_H}$)
for the model with an $A_H$ particle (dashed lines) and the model with
no $A_H$ particle (solid lines).
Left: Precision on $f$ as a function of $x$, for $f = 1$, 2, 3, 4 TeV
(bottom to top for each line type).
Right: Precision on $f$ as a function of $f$, for $x=0.37$.  
}
\label{fig:f}
\end{figure}
The strongest $f$ dependence (and thus the highest precision desired) 
occurs for $x \simeq 0.37$, as can be seen in the left panel of 
Fig.~\ref{fig:f}.  
In the right panel of Fig.~\ref{fig:f} we show the precision with which 
$f$ must be measured as a function of $f$, taking $x=0.37$ to conservatively
give the strongest $f$ dependence.  The electroweak precision data 
constrain the scale $f$ to be no smaller than about 1 TeV
\cite{GrahamEW2}.  From the right panel of Fig.~\ref{fig:f}, 
$f = 1$ TeV corresponds to a required precision of 
$\delta f/f \leq 7\%$.  For $f > 3.5$ TeV, the precision $\delta f/f$ 
required to give $\delta R/R_{\rm SM} = 1\%$ is greater than one, meaning 
that knowing that $0 < f < 7$ TeV is sufficient.
However, for such
high $f$ values, the correction to the rate for 
$\gamma\gamma \to H \to b \bar b$ due to the Littlest Higgs model is 
comparable in size to the $1\sigma$ experimental resolution \cite{LHloop}, 
and the measurement loses its usefulness as a test of the model.

In the first parameterization, the $f$ dependence is slightly stronger
than that shown in Fig.~\ref{fig:f}.
This drives our choice of the input parameter set: by choosing to work in the
second parameterization, we reduce the precision with which $f$ must
be determined.  In addition, we trade two mixing angles, $c$ and $c^{\prime}$,
whose values must be extracted from a combination of measurements, for
the masses of two heavy gauge bosons, $M_{Z_H}$ and $M_{A_H}$, which
can be measured directly.\footnote{A full analysis would compute the
rate for $\gamma\gamma \to H \to b \bar b$ from a fit of the model 
parameters based on all LHC data, in which case choosing a 
parameterization would be unnecessary.  Such a fit is beyond the scope
of our current work, which seeks only to estimate whether the parameter
uncertainties from the LHC measurements will be small enough to give a
reliable prediction for the rate for $\gamma\gamma \to H \to b \bar b$ in
the Littlest Higgs model.}

How can $f$ be measured at the LHC?  The most obvious approach is to
extract $f$ from the measurements of the $Z_H$ mass and cross
section.  The $Z_H$ mass depends on $f$ and $c$ as given in 
Eq.~(\ref{eq:heavygaugemasses}).
The $Z_H$ will most likely be discovered in Drell-Yan production with
decays to $e^+e^-$ or $\mu^+\mu^-$.  For fixed $M_{Z_H}$, the rate for 
$pp \to Z_H \to \ell^+\ell^-$ depends strongly on the parameter $c$ 
through both the production cross section (proportional to $\cot^2\theta$)
and the decay branching ratio of $Z_H$ to dileptons \cite{Burdman,LHpheno}.
Neglecting the masses of the final-state particles
compared to $M_{Z_H}$, the $Z_H$ partial width into a pair of fermions 
is given by 
\begin{equation}
        \Gamma (Z_H \to f \bar f) = \frac{N_c g^2 \cot^2 \theta}{96 \pi}
        M_{Z_H},
\end{equation}
where $N_c$ = 3 for quarks and 1 for leptons, and the partial width into
boson pairs is given by
\begin{equation}
        \Gamma (Z_H \to Z H) = \Gamma (Z_H \to WW)
        = \frac{g^2 \cot^2 2\theta}{192 \pi} M_{Z_H}.
\end{equation}
In our numerical calculations of $Z_H$ branching fractions we ignore
the masses of all final-state particles except for the top quark.

\begin{figure}
\resizebox{0.5\textwidth}{!}{\rotatebox{270}{
\includegraphics[50,50][555,600]{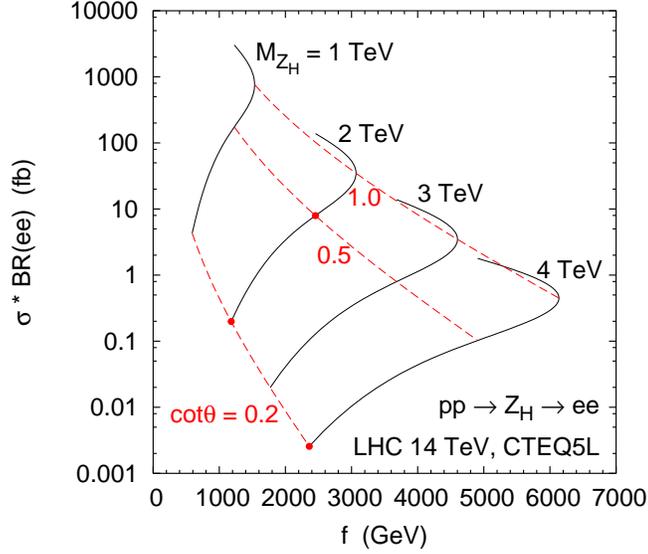}}}
\caption{Cross section times branching ratio for $Z_H$ into dielectrons
at the LHC as a function of $f$.  The solid black lines are contours of
constant $M_{Z_H}$, while the dashed red lines are contours of constant
$\cot\theta$.}
\label{fig:sigmatimesBR2}
\end{figure}

In Fig.~\ref{fig:sigmatimesBR2} we show the cross section for $Z_H$ 
times its branching ratio into dielectrons as a function of $f$.
Electroweak precision data requires $f \gsim 1$ TeV and $M_{Z_H} \gsim 2$ TeV
\cite{Kilian,GrahamEW1,JoAnneEW,GrahamEW2}.  
Perturbativity of the two SU(2) gauge couplings,
$g_{1,2} \lsim \sqrt{4\pi}$, requires $\cot\theta \gsim 0.18$.
With these constraints, a wide range of cross sections are allowed.

\begin{figure}
\resizebox{\textwidth}{!}{\rotatebox{270}{
\includegraphics[50,50][555,600]{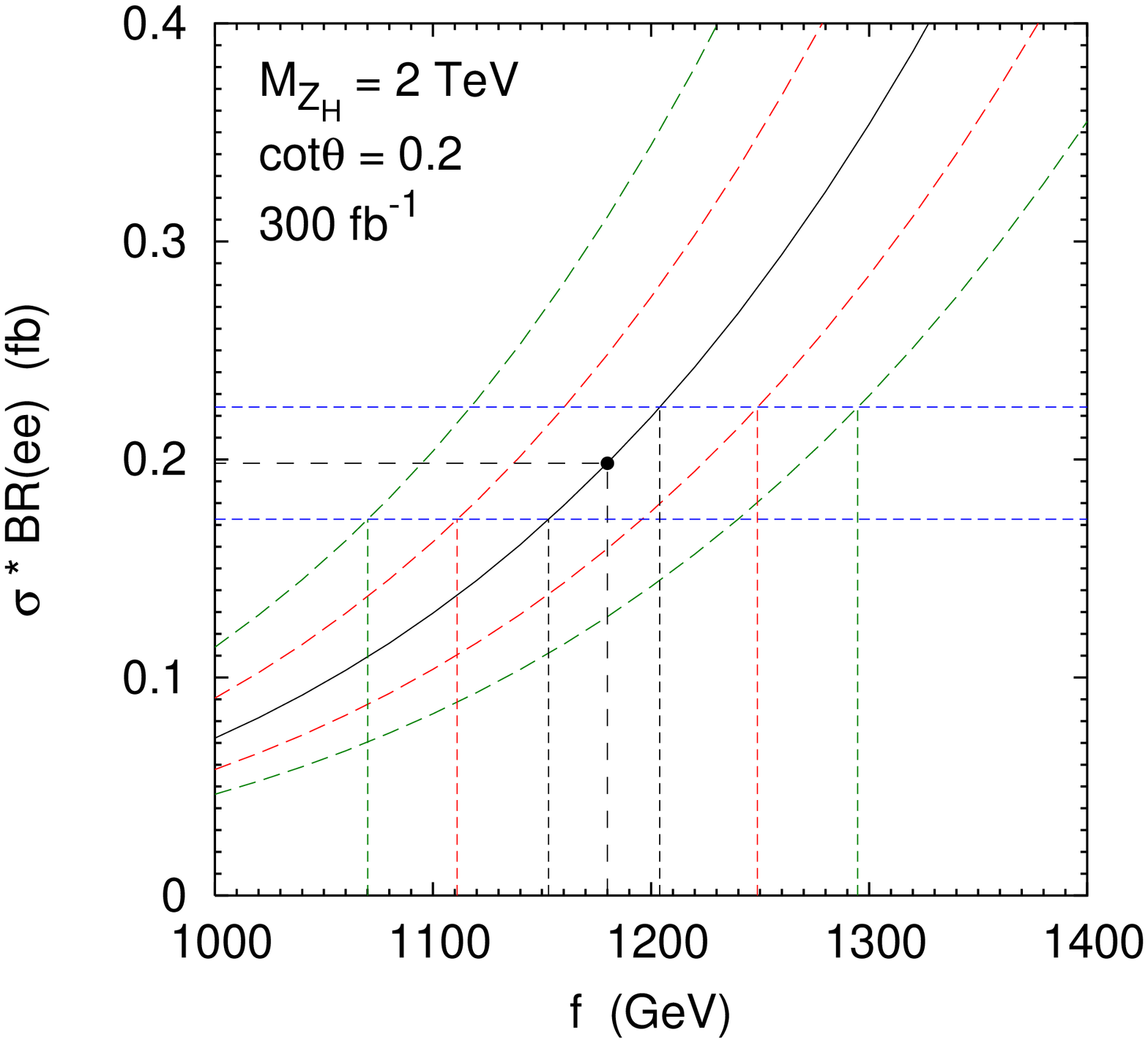}}
\rotatebox{270}{
\includegraphics[50,50][555,600]{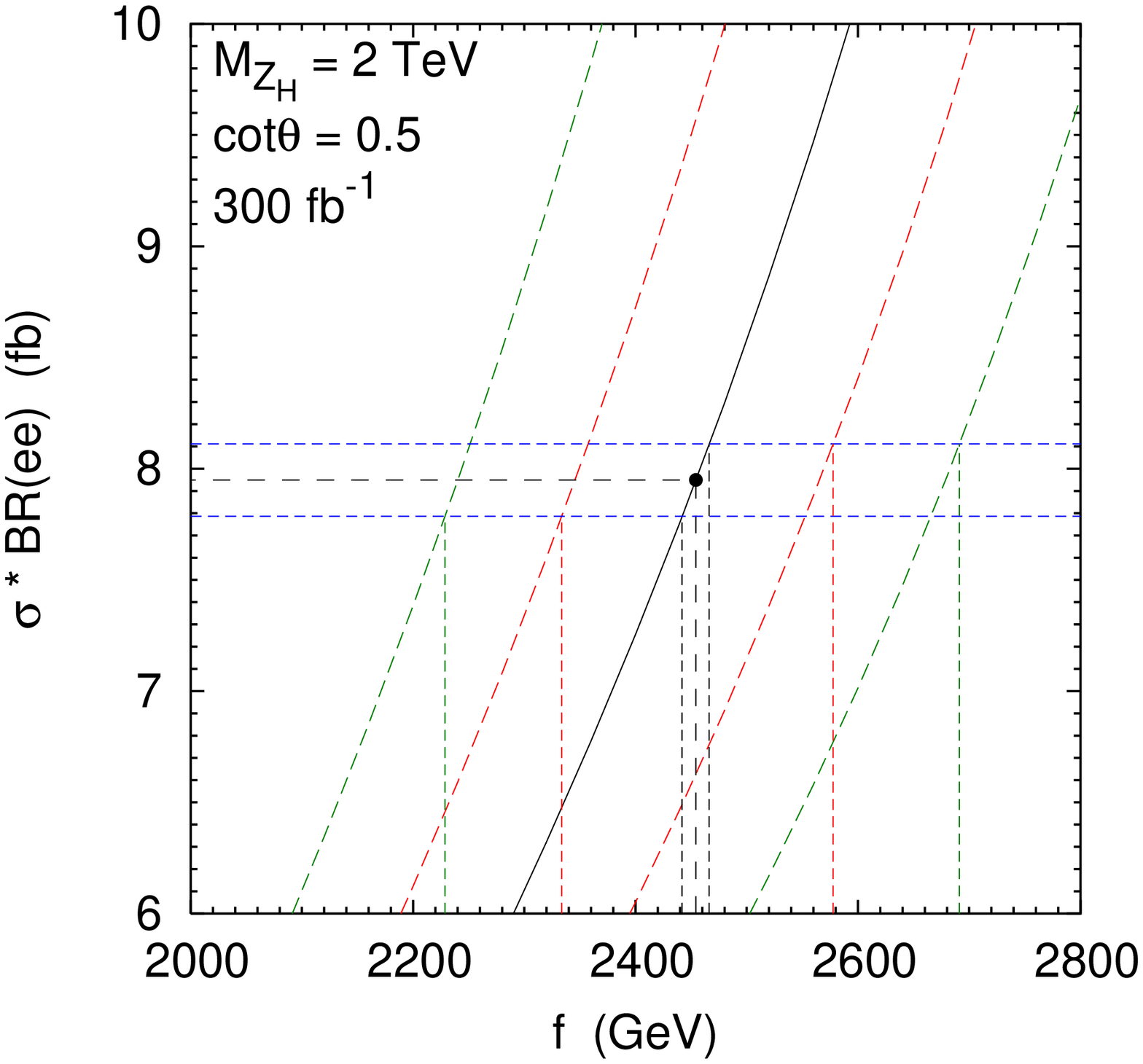}}}
\caption{Cross section times branching ratio for $Z_H$ into dielectrons
at the LHC for Points 1 (left) and 2 (right) discussed in the text (black
dots).  The solid black line is the contour of $M_{Z_H} = 2$ TeV.  The
dashed red and green lines are contours of $M_{Z_H} = 2$ TeV $\pm 2\%$ and 
4\%, respectively, and show the effect of the finite $M_{Z_H}$ mass 
resolution on the $f$ determination.  
The horizontal short-dashed blue lines show the $1\sigma$
statistical uncertainty in the cross section, assuming 100\% acceptance
and 300 fb$^{-1}$ of data.}
\label{fig:fscale-points}
\end{figure}

A measurement of the $Z_H$ cross section times its branching ratio into
dielectrons (from counting events) can be combined with a measurement 
of $M_{Z_H}$ (from the dielectron invariant mass) to extract $f$.
To illustrate the prospects for measuring $f$, we study three benchmark
points:
\begin{itemize}
\item Point 1: $M_{Z_H} = 2$ TeV, $\cot\theta = 0.2$, corresponding to
$f = 1180$ GeV;
\item Point 2: $M_{Z_H} = 2$ TeV, $\cot\theta = 0.5$, corresponding to
$f = 2454$ GeV;
\item Point 3: $M_{Z_H} = 4$ TeV, $\cot\theta = 0.2$, corresponding to
$f = 2360$ GeV.
\end{itemize}
The $f$ extraction from the cross section measurement is illustrated
for Points 1 and 2 in Fig.~\ref{fig:fscale-points}.
The resulting uncertainty $\delta f/f$ is summarized in 
Table~\ref{tab:fscale-points}.
\begin{table}
\begin{tabular}{l|c|c|ccc|c}
\hline \hline
        &       & Statistical uncertainty 
        & \multicolumn{3}{c|}{$\delta f/f$} & Desired $\delta f/f$ \\
     & $f$ (GeV) & on $\sigma \times {\rm BR}(ee)$ &
        ($\delta M_{Z_H} = 0$) &
        ($2\%$) &
        ($4\%$) &
        (no $A_H$/with $A_H$) \\
\hline
Point 1 & 1180 & 13\% (59 evts) & 2\% & 6\% & 10\% & 10\% / 12\% \\ 
Point 2 & 2454 & 2.0\% (2380 evts) & 0.5\% & 5\% & 9\% & 43\% / 49\% \\
Point 3 & 2360 & -- (0.8 evts) & -- & -- & -- & 40\% / 45\% \\
\hline \hline
\end{tabular}
\caption{Extraction of $f$ from the $Z_H$ mass and rate in dielectrons
for the three points discussed in the text.  The statistical uncertainty
on the cross section times branching ratio is calculated from the 
number of dielectron events assuming 100\% acceptance and 
300 fb$^{-1}$ of data.  
The effect of the $M_{Z_H}$ measurement uncertainty is also shown for
$\delta M_{Z_H}/M_{Z_H} = 0$, 2\% and 4\%.
The desired $\delta f/f$ is taken from Fig.~\ref{fig:f} for the 
versions of the model without and with an $A_H$ boson.}
\label{tab:fscale-points}
\end{table}
It is possible to achieve the desired precision on $f$ to give 
$\delta R/R_{\rm SM} = 1\%$ (Fig.~\ref{fig:f}) over a large part of 
the parameter space.  For Points 1 and 2, the uncertainty in the 
$M_{Z_H}$ measurement dominates the uncertainty in $f$ for 
$\delta M_{Z_H} \gsim 2\%$.  To match the desired precision for the low
$f \sim 1.2$ TeV of Point 1, a fairly high precision measurement of 
the $Z_H$ mass, $\delta M_{Z_H}/M_{Z_H} \sim 4\%$, is required.
Point 3 was chosen as a worst-case scenario with very small cross section
yet a moderate value of $f \sim 2.3$ TeV.  At this parameter point $Z_H$ 
will not be detected at the LHC in dileptons since the number of events 
is too small.  The bosonic decay modes have larger branching fractions
at this point \cite{Burdman,LHpheno}, but the $Z_H$ is still 
unlikely to be detected
in the bosonic channels for the parameters of Point 3 \cite{ATLASLH}.

In Fig.~\ref{fig:fscale-points} and Table~\ref{tab:fscale-points}
the statistical uncertainty on the cross section times branching ratio
is taken as $\sqrt{N_S}$ for $N_S$ signal events.  The number of signal
events we take to be $\sigma \times {\rm BR}(ee) \times 300$ fb$^{-1}$;
that is, we assume 100\% acceptance for dielectron events in the $Z_H$
mass window on top of negligible background.  This is of course optimistic;
however, very minimal cuts should be needed for the $Z_H$ reconstruction in 
dileptons.  The statistics used in Fig.~\ref{fig:fscale-points}
and Table~\ref{tab:fscale-points}
can be doubled by including the dimuon channel,
and doubled again by including data from both of the two LHC detectors.  

\begin{figure}
\resizebox{\textwidth}{!}{\rotatebox{270}{
\includegraphics[50,50][555,600]{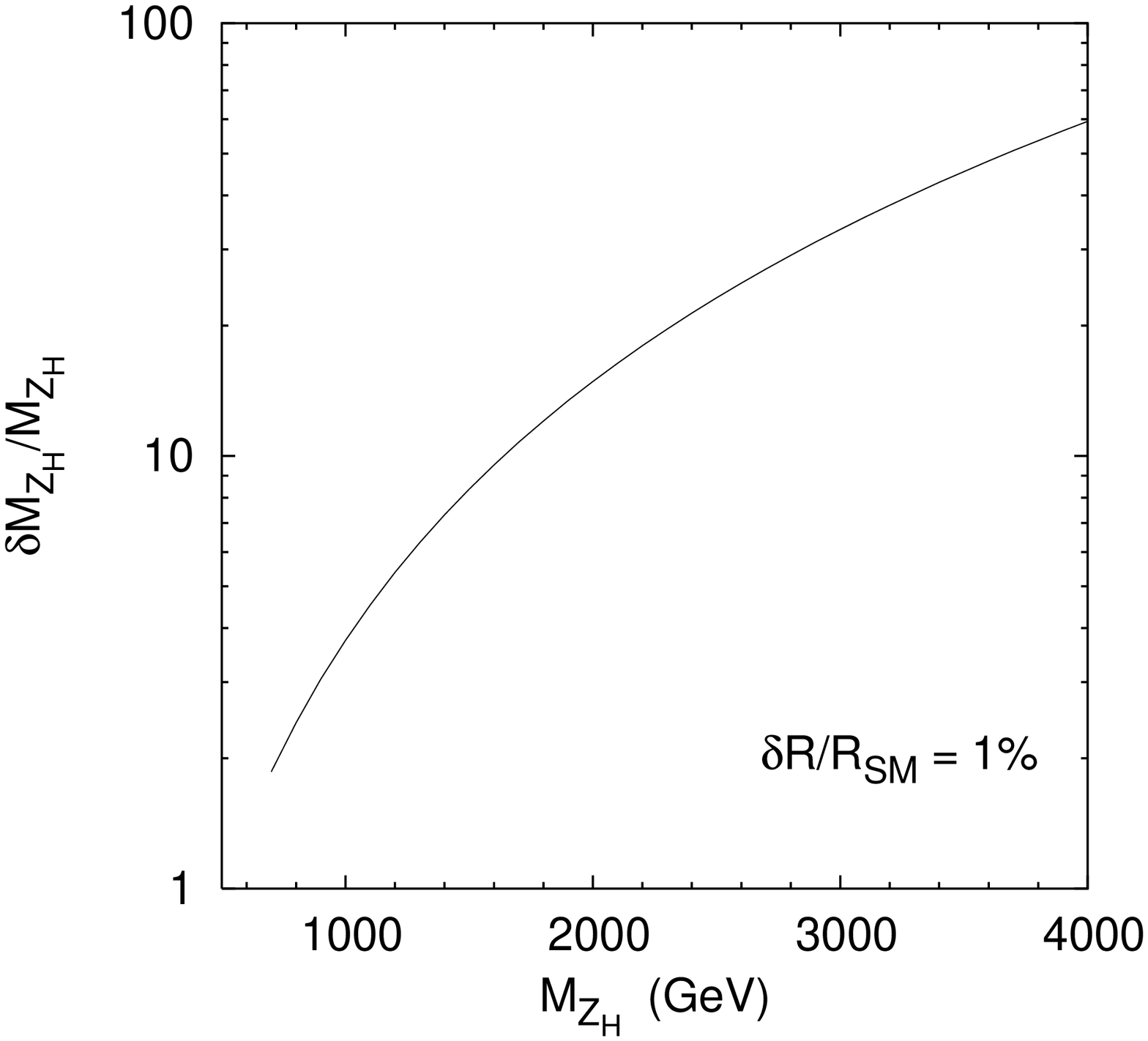}}
\rotatebox{270}{
\includegraphics[50,50][555,600]{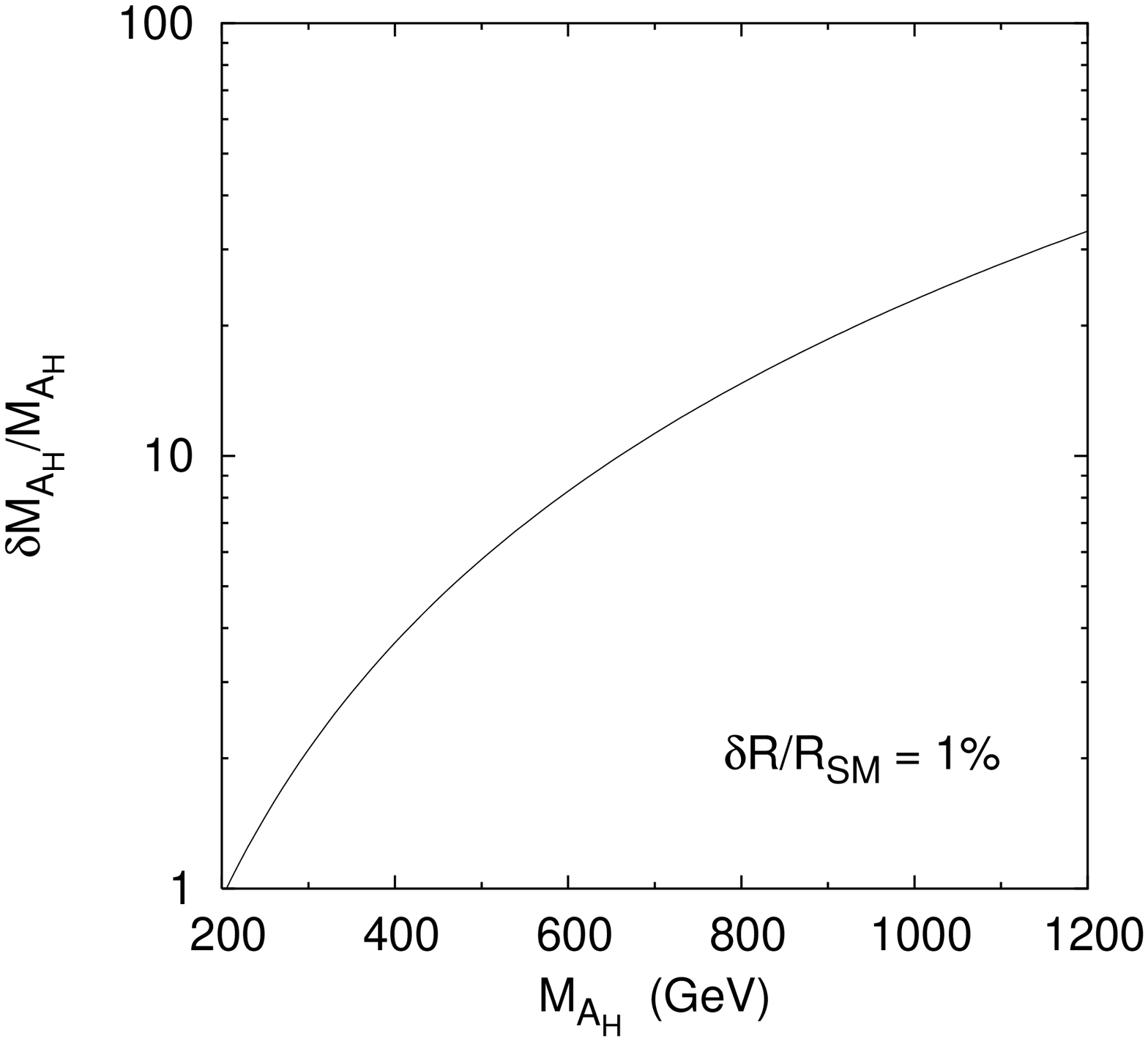}}}
\caption{Precision on $M_{Z_H}$ (left) and $M_{A_H}$ (right) required
for $\delta R/R_{\rm SM} = 1\%$.}
\label{fig:MZH}
\end{figure}

Finally we consider the masses of the heavy gauge bosons $Z_H$ and $A_H$,
shown in Fig.~\ref{fig:MZH}.
Because the $M_{Z_H}$ ($M_{A_H}$) dependence of the corrections to the
Higgs couplings can be separated from that of the other parameters,
the precision needed on $M_{Z_H}$ ($M_{A_H}$) is independent of the
other parameter values.

The electroweak precision data constrain the masses of the heavy SU(2)
gauge bosons $Z_H$, $W_H^{\pm}$ to be no lighter than about 2 TeV
\cite{GrahamEW1,JoAnneEW,GrahamEW2}.  
From the left panel of Fig.~\ref{fig:MZH}, the precision 
$\delta M_{Z_H}/M_{Z_H}$ required to give $\delta R/R_{\rm SM} = 1\%$ 
is greater than one, meaning that only a very rough knowledge
of this parameter is required.  In particular, even for
$M_{Z_H} = 1$ TeV, $M_{Z_H}$ need only be known within a factor
of three.  This precision will be trivial to achieve.  The requirement on 
the $Z_H$ mass measurement for the extraction of $f$ is much more stringent.

If the model contains an $A_H$ gauge boson, a measurement of its mass
will only be important if it is lighter than about 200 GeV.  For a
heavier $A_H$, the precision $\delta M_{A_H}/M_{A_H}$ required to give
$\delta R/R_{\rm SM} = 1\%$ is greater than one (right panel of 
Fig.~\ref{fig:MZH}).

\section{Other uncertainties}
\label{sec:uncert}

In addition to the parametric uncertainties in the calculation of the
rate for $\gamma\gamma \to H \to b \bar b$, we must consider other 
sources of uncertainty.  In this section we discuss potential 
theoretical and experimental uncertainties in the Littlest Higgs 
model parameter extraction, issues in the extraction of 
$\Gamma_{\gamma} \times {\rm BR}(H \to b \bar b)$
from photon collider measurements,
and the sources of uncertainty in the SM Higgs coupling calculation.

\subsection{Littlest Higgs parameter extraction}

We have computed the correction to the Higgs partial widths working to 
leading nontrivial order in the expansion of the Littlest Higgs nonlinear 
sigma model in powers of $v^2/f^2$.  Higher-order corrections to the
Higgs partial widths from the $v^2/f^2$ expansion are unlikely to be 
relevant.  Higher-order corrections to the parameter translations
(e.g., $f,c \leftrightarrow M_{Z_H}$) and parameter extractions from LHC
data, however, could be important.  Their effects on the parameter 
extraction will be at the few-percent level, which is relevant in 
particular for the $Z_H$ mass in the extraction of $f$ at low $f$ values.
These higher-order terms in the expansion are straightforward to include.

QCD corrections to the cross section for $Z_H$ production at the LHC 
must be taken into account in the determination of the $f$ scale.
These can be taken over directly from the SM computations for 
$Z, \gamma$-mediated Drell-Yan.
The next-to-leading order (NLO) QCD corrections to Drell-Yan were 
computed some 25 years ago \cite{DYNLO} and yield $K$-factors of 
order 1.4.  With the computation of the inclusive NNLO $K$-factor
more than ten years ago \cite{DYNNLO} and the recent computation of 
the differential NNLO cross section within the past year 
\cite{DYNNLOdiff}, the QCD uncertainty in the $Z_H$ cross section 
is well under control.  Similarly, the (relatively small) QCD 
corrections to the $Z_H$ branching fraction to dileptons can be 
taken over from the corresponding calculation for $Z$ decays.
In addition, the LHC luminosity uncertainty will contribute to the
uncertainty in the $Z_H$ cross section.
However,
a quick examination of Table~\ref{tab:fscale-points} reveals that 
even a $\sim 10\%$ (statistical) uncertainty on the $Z_H$ production 
cross section times leptonic branching ratio does not contribute
significantly to the uncertainty in $f$, so that these systematic
uncertainties are not a problem.

More important for the determination of $f$ are the corrections to, 
and measurement uncertainties of, the $Z_H$ boson mass.  For 
$f \sim 1$ TeV, a measurement of $M_{Z_H}$ at the $\sim 4\%$ level is 
desirable.  At this level of precision, electroweak radiative corrections
to the $Z_H$ mass could be important.  To be more precise, the parameter
translations between LHC measurements, Littlest Higgs model parameters,
and the $\gamma\gamma \to H \to b \bar b$ rate may need to be treated
at next-to-leading order in the electroweak couplings.  This could also be
important for the parameter $x$ at low $f \sim 1$ TeV, which we have 
proposed to
extract from the $W$ boson mass measurement.  Radiative corrections to
the $W$ mass within the Littlest Higgs model could be important for 
this extraction; in particular, because the model contains a scalar 
triplet that gets a nonzero vev, violating custodial symmetry at the 
tree level, the renormalization of the electroweak sector at the 
one-loop level requires one additional input (to fix the triplet vev
counterterm) beyond the usual three SM inputs \cite{phivev,Sally}.
This extra input parameter can have important effects on the parameter
dependence of the one-loop corrections to the SM observables 
\cite{Sally,phivev2}.\footnote{We thank Sally Dawson for pointing out
this complication.}  The first one-loop calculation in the Littlest Higgs
model involving renormalization of the electroweak sector was done
in Ref.~\cite{Sally}.

There are also experimental issues in the measurement of the $Z_H$ 
mass to high precision, which have been discussed, e.g., in 
Ref.~\cite{LesHouches}.  
The measurement of the mass of a new heavy
(TeV-scale) gauge boson at the LHC relies on accurate measurements of
the energy/momentum of very high-energy electrons or muons.  For the 
$Z_H$ masses considered ($\geq 2$ TeV), these leptons will have energies 
of 1 TeV or higher.  For electrons, the energy measurement will come
primarily from the electromagnetic calorimeter.  Uncertainties come from
both the energy resolution and the energy scale calibration.
A calibration of the lepton energy scale at TeV-scale energies could be 
made, e.g., using very high-$p_T$ $Z$ bosons decaying to dielectrons.
For muons, the momentum is measured from track curvature.  While the
calibration is under control here, the energy resolution per event is 
worse because the tracks are very stiff, so higher statistics may be 
needed.  Since many models of TeV-scale new physics contain high-mass
resonances that decay to dileptons, we feel that a more detailed study
of the systematic uncertainties affecting the $Z^{\prime}$ mass 
measurement would be worthwhile.

\subsection{Photon collider issues}

Photon collider studies 
\cite{ggAsner,LeptonPhoton,AGG,Jikia,Krawczyk1,Krawczyk2,Rosca}
claim a 2\% measurement of the $\gamma\gamma \to H \to b \bar b$ rate,
which we interpret as a measurement of 
$\Gamma_{\gamma} \times {\rm BR}(H \to b \bar b)$.  We mention here 
some sources of uncertainty that must be under control before such a 
high precision measurement is claimed.

First, the $\gamma\gamma$ luminosity and polarization spectra must be 
measured to normalize the Higgs production rate.  
The photon and electron luminosity and polarization spectra are currently 
simulated using the programs CAIN \cite{CAIN} and GUINEA-PIG 
\cite{GUINEA-PIG}.  The luminosity spectrum 
can be measured using the reactions $\gamma\gamma \to e^+e^-$ 
\cite{gglumipol} and perhaps $\gamma\gamma \to e^+e^- \gamma$.  The photon 
polarization spectrum could be measured using $e\gamma \to e\gamma$ and 
$e\gamma \to W\nu$ \cite{TESLATDR,ggAsner}.\footnote{The TESLA Conceptual
Design \cite{TESLACD} considered a scheme in which the spent electrons
were deflected away from the interaction region using magnets; however,
the current photon collider designs do not include magnetic deflection.}  
Further study is needed.

Second, a photon collider collides more than just photons.  The photon
has a parton distribution function containing quarks, gluons, etc., and
collisions of such ``resolved'' photons can yield Higgs production via,
e.g., gluon fusion or $b \bar b$ fusion.  This resolved-photon part of 
the Higgs production cross section is not proportional to $\Gamma_{\gamma}$.
The resolved photon contribution to SM Higgs production has been studied
in Ref.~\cite{Doncheski} for a photon collider with 
$\sqrt{s}_{ee} = 500$ GeV and found to be at the percent level or 
smaller.\footnote{Resolved photon contributions to the 
background $b \bar b$ production
were studied in Ref.~\cite{resolvedphotonbg} and found to be small
if the photon collider beam energy is optimized for Higgs production.}
Similarly, the remnant electron beams can contribute to Higgs production 
via $ZZ$ fusion, $e^-e^- \to e^-e^- Z^*Z^* \to e^-e^- H$.
These contributions to Higgs production are likely to be small, but 
a quantitative estimate would be useful.

Finally, the background to $\gamma\gamma \to H \to b \bar b$ consists 
mostly of $b \bar b (g)$ production, with some $c \bar c (g)$ contribution
from charm quarks mistagged as bottom.  The signal is peaked at 
the Higgs mass on top of a background steeply falling with increasing 
two-jet invariant mass (due to the photon beam energy spectrum).  The 
background can be simulated based on the beam spectra \cite{CAIN,GUINEA-PIG}
and the QCD-corrected cross sections for heavy quark pair production
in $\gamma\gamma$ collisions \cite{ggtobbg}.
The background normalization must be under control to subtract from the 
signal.

\subsection{Standard Model Higgs coupling calculation}

In order to predict the rate for $\gamma\gamma \to H \to b \bar b$
at the 1\% level in the Littlest Higgs model, the SM rate must be
known at the same level of precision.  We outline here the known 
radiative corrections and sources of uncertainty in the SM prediction.

The SM $H \to \gamma\gamma$ decay partial width receives QCD corrections, 
which of course only affect the top-quark diagrams.  Because the 
external particles in the $\gamma\gamma H$ vertex are color neutral, 
the virtual QCD corrections are finite by themselves.  Since no real
radiation diagrams contribute, the QCD corrections to $H \to \gamma\gamma$
are equivalent to those to the inverse process $\gamma\gamma \to H$.  
This is in contrast to, e.g., the QCD corrections to the $ggH$ vertex.

The QCD corrections to $\Gamma_{\gamma}$ in the SM 
are known analytically at the 
two-loop [$\mathcal{O}(\alpha_s)$] order \cite{ggH2loopQCD} 
and as a power expansion up to third order in $M_H/m_t$ at three-loop 
[$\mathcal{O}(\alpha_s^2)$] order \cite{ggH3loopQCD}.  They are small
for Higgs masses $M_H < 2 m_t$; the $\mathcal{O}(\alpha_s)$ corrections
are only of order 2\% for $M_H < 2 M_W$, and the $\mathcal{O}(\alpha_s^2)$
corrections are negligible, demonstrating that the QCD corrections are 
well under control.

The SM $H \to \gamma\gamma$ decay partial width also receives electroweak
radiative corrections.  The the electroweak corrections are much more 
difficult to compute than the QCD corrections and a full two-loop calculation
does not yet exist.  The electroweak correction due to two-loop diagrams
containing light fermion loops and $W$ or $Z$ bosons 
(with the Higgs boson coupled to the $W$ or $Z$ boson, because the light 
fermion Yukawa couplings are neglected) was computed recently in 
Ref.~\cite{ggHEWlightf} and contributes between $-1\%$ and $-2\%$ for 
$M_H \lsim 140$ GeV.
The leading $\mathcal{O}(G_F m_t^2)$ electroweak correction due to 
top-mass-enhanced two-loop diagrams containing third-generation 
quarks was also computed recently in Ref.~\cite{ggHEWGFmt2} as an expansion
to fourth order in the ratio $M_H^2/(2M_W)^2$.\footnote{The 
$\mathcal{O}(G_F m_t^2)$ electroweak correction was also considered in
Ref.~\cite{ggHEWGFmt2old}, whose results disagree with that of 
Ref.~\cite{ggHEWGFmt2}.  The source of this disagreement is addressed
in Ref.~\cite{ggHEWGFmt2}.}  The expansion appears to
be under good control for $M_H \lsim 140$ GeV, where this correction 
contributes about $-2.5\%$ almost independent of $M_H$.\footnote{The 
leading $\mathcal{O}(G_F M_H^2)$ correction was computed
in Ref.~\cite{ggHEWGFMH2} for large $M_H$; however, this limit is
not useful for the light Higgs boson that we consider here.}
We conclude that the electroweak radiative corrections to 
$\gamma\gamma \to H$ appear to be under control at the 1--2\% level.

We now consider the uncertainty in the SM prediction for the $H \to b \bar b$
branching ratio.  
The radiative corrections to Higgs decays to fermion and boson pairs
have been reviewed in Ref.~\cite{Spirareview}; we give here a brief
sketch of the known corrections and refer to Ref.~\cite{Spirareview}
for references to the original calculations.
The full QCD corrections to the Higgs decay to $q \bar q$ are known
up to three loops neglecting the quark mass in the kinematics and 
up to two loops for massive final-state quarks.
The electroweak corrections to the Higgs decay to quark
or lepton pairs are known at one-loop; in addition, the QCD corrections 
to the leading top-mass-enhanced electroweak correction term are known up
to three loops, to order $G_Fm_t^2 \alpha_s^2$.
All of these corrections to the Higgs partial widths to fermions
are included in a consistent way in the program HDECAY \cite{HDECAY}.

For the Higgs masses below the $WW$ threshold that we consider here,
decays into off-shell gauge bosons ($WW$, $ZZ$) are important and 
affect the total Higgs width, thus feeding in to ${\rm BR}(H \to b \bar b)$.
HDECAY takes into account decays with both $W$ ($Z$) bosons
off-shell.  One-loop electroweak corrections to Higgs decays to $WW$
and $ZZ$ are known, together with the QCD corrections
to the leading $\mathcal{O}(G_F m_t^2)$ result up to three loops.
These corrections to $\Gamma_{W,Z}$ amount to less 
than about 5\% in the intermediate Higgs mass range \cite{Spirareview}
(translating to less than roughly 2\% in ${\rm BR}(H \to b \bar b)$
for $M_H \simeq 120$ GeV) and have been neglected in HDECAY, although
their inclusion would seem straightforward.

The $H \to b \bar b$ branching ratio in the SM also has a parametric 
uncertainty due to the nonzero present experimental uncertainties in the 
SM input parameters.  The largest sources of parametric uncertainty
are the bottom quark mass and (to a lesser extent) the strong coupling 
$\alpha_s$ (which 
contributes via the QCD corrections to the $H b \bar b$ coupling).  
This parametric uncertainty in ${\rm BR}(H \to b \bar b)$ was evaluated 
in Ref.~\cite{MSSMHiggs} to be about 1.4\% for $M_H = 120$ GeV, using 
the standard $\alpha_s = 0.1185 \pm 0.0020$ \cite{alphas} and 
a somewhat optimistic $m_b(m_b) = 4.17 \pm 0.05$ GeV 
($\overline{\rm MS}$) \cite{mbmb}.
The parametric uncertainty in the branching ratio is suppressed due to 
the fact that $\Gamma_b$ makes up about 2/3 of the Higgs total
width at $M_H = 120$ GeV, leading to a partial cancellation of the 
uncertainty in the branching ratio; we thus expect the parametric 
uncertainty to be somewhat larger at higher Higgs masses, where 
$\Gamma_b$ no longer dominates the total width.

The best measurements of $\alpha_s$ come from LEP-I and II; the Tevatron
and LHC are unlikely to improve on this.
The bottom quark mass is extracted from heavy quarkonium spectroscopy 
and $B$ meson decays with a precision limited by theoretical uncertainty.
There are prospects to improve the bottom quark mass extraction through
better perturbative and lattice calculations \cite{bottommass}
and more precise measurements of the upsilon meson properties from CLEO
\cite{CLEO}.

\section{Conclusions}
\label{sec:conclusions}

We have calculated the $\mathcal{O}(v^2/f^2)$ corrections to the partial 
widths of the light Higgs boson in the Littlest Higgs model.  These
results allow numerical calculations of the corrections to the Higgs
boson total width and decay branching ratios, as well as the corrections
to the Higgs boson production cross section in two-photon fusion and
in gluon fusion.  We studied the correction to the rate of 
$\gamma\gamma \to H \to b \bar b$, which is expected to be measured
at a future photon collider with 2\% precision for a light Higgs boson 
with mass in the range $115-140$ GeV.

For $f \sim 1$ TeV, the correction to the $\gamma\gamma \to H \to b \bar b$
rate is roughly $-6\%$.  In order to make a theoretical
prediction for the corrected rate $R = R_{\rm SM} + R_{\rm LH}$ with 1\%
precision (i.e., a theoretical uncertainty comfortably 
smaller than the experimental
uncertainty of 2\%), the correction $R_{\rm LH}$ need only be computed
at the $15\%$ level for $f \sim 1$ TeV.  We studied the precision with
which the Littlest Higgs model parameters must be measured in order to
match the photon collider precision, and conclude that measurements of 
the model parameters with high enough precision should be possible at 
the LHC over much of the relevant model parameter space.

The measurement of $\gamma\gamma \to H \to b \bar b$ provides a 
nontrivial test of the Littlest Higgs model.  More interestingly, 
it also provides a probe
of the UV completion of the nonlinear sigma model at the 10 TeV scale.
The loop-induced Higgs coupling to photon pairs, for example,
can receive corrections from the new heavy particles of the UV completion
running in the loop.  Equivalently, the dimension-6 operator
$h^{\dagger} h F^{\mu\nu} F_{\mu\nu} / \Lambda^2$ that gives rise to
the $\gamma\gamma H$ coupling receives a contribution from the 10 TeV scale.  
If the UV completion is weakly coupled, these corrections will be
suppressed by the square of the ratio of the electroweak scale to 
the 10 TeV scale, and thus be too small to detect with the expected 
2\% experimental resolution.  If the UV completion is strongly coupled, 
however, the strong-coupling enhancement counteracts the suppression from
the high mass scale, leading to corrections parametrically of the same 
order as those from the TeV scale physics that should be observable
at the photon collider.

\begin{acknowledgments}
We thank Jack Gunion, Tao Han, Bob McElrath, Mayda Velasco, 
and Lian-Tao Wang for valuable discussions, and Frank Paige 
for elucidating the lepton energy measurement issues at the LHC.
We also thank the organizers of the ALCPG 2004 Winter Workshop at SLAC
where preliminary results were presented.
This work was supported in part by the U.S.~Department of Energy
under grant DE-FG02-95ER40896
and in part by the Wisconsin Alumni Research Foundation.
\end{acknowledgments}

\appendix
\section{}

The partial width of the Higgs boson into two photons is 
given in the Littlest Higgs model by \cite{LHloop,HHG}
\begin{equation}
        \Gamma_\gamma = 
        \frac{\sqrt{2} G_F \alpha^2 M_H^3 y^2_{G_F}}{256 \pi^3}
        \left| \sum_i y_i N_{ci} Q_i^2 F_i \right|^2,
	\label{eq:Gga}
\end{equation}
where $N_{ci}$ and $Q_i$ are the color factor ($=1$ or 3) and electric charge, 
respectively, for each particle $i$ running in the loop.
The standard dimensionless loop factors
$F_i$ for particles of spin 1, 1/2, and 0 are given in Ref.~\cite{HHG}.
The factors $y_i$ in the sum incorporate the couplings and mass suppression
factors of the particles running in the loop.  For the top quark and 
$W$ boson, whose couplings to the Higgs boson are proportional to their masses,
the $y_i$ factors are equal to one up to a correction of order 
$v^2/f^2$ \cite{LHloop}.
For the TeV-scale particles in the loop, on the other hand, the $y_i$ 
factors are of order $v^2/f^2$.  This reflects the fact that the masses
of the heavy particles are not generated by their couplings to the 
Higgs boson; rather, they are generated by the $f$ condensate.
This behavior naturally respects the 
decoupling limit for physics at the scale $f \gg v$.

Normalizing the Higgs partial width into photons to its SM value,
we have
\begin{equation}
        \Gamma_{\gamma} / \Gamma_{\gamma}^{\rm SM}
        = y^2_{G_F} \frac{\left| \sum_{i,{\rm LH}} 
                y_i N_{ci} Q_i^2 F_i \right|^2}
        {\left| \sum_{i,{\rm SM}} N_{ci} Q_i^2 F_i \right|^2},
        \label{eq:Gga/SM}
\end{equation}
where $i$ runs over the fermions in the loop: 
$t$, $T$, $W$, $W_H$, and $\Phi^+$ in the Littlest Higgs (LH) case; 
and $t$ and $W$ in the SM case.

The partial width of the Higgs boson into two gluons, normalized to its
SM value, is given in the Littlest Higgs model by \cite{HHG,LHloop}
\begin{equation}
        \Gamma_g / \Gamma_g^{\rm SM} = y^2_{G_F}
        \frac{ \left| \sum_{i,{\rm LH}} y_i F_{1/2}(\tau_i) \right|^2 }
        { \left| \sum_{i,{\rm SM}} F_{1/2}(\tau_i) \right|^2 },
	\label{eq:Gg/SM}
\end{equation}
where $i$ runs over the fermions in the loop: 
$t$ and $T$ in the Littlest Higgs case, and $t$ in the SM case.
The dimensionless loop factor $F_{1/2}$ is again given in Ref.~\cite{HHG}.

We now list the formulas for the correction factors $y_i$
in terms of two sets of input parameters:
\begin{enumerate}
\item $c_t$, $x$, $f$, $c$, $c^{\prime}$, and
\item $c_t$, $x$, $f$, $M_{Z_H}$, $M_{A_H}$.
\end{enumerate}

For the model in which two U(1) groups are gauged, leading to 
an $A_H$ particle in the spectrum, we have:\footnote{We thank 
J\"urgen Reuter for correspondence leading to the correction of 
errors in Eqs.~(\ref{eq:yt})-(\ref{eq:yT}) and (\ref{eq:y2cW}) 
in an earlier version of this manuscript.} 
\begin{eqnarray}
        y^2_{G_F} &=& 1 + \frac{v^2}{f^2} 
                \left[ -\frac{5}{12} + \frac{1}{4} x^2 \right]
                \label{eq:y2GF}
        \\
        y_t &=& 1 + \frac{v^2}{f^2} \left[ -\frac{2}{3} + \frac{1}{2} x
                - \frac{1}{4} x^2 + c_t^2 s_t^2 \right] 
	\label{eq:yt}
        \\
        y_W &=& 1 + \frac{v^2}{f^2} \left[ -\frac{1}{6} 
                - \frac{1}{4} (c^2-s^2)^2 \right]
                \nonumber \\
                &=& 1 + \frac{v^2}{f^2} \left[ -\frac{5}{12} \right]
                + \frac{M_W^2}{M_{Z_H}^2}
        \\
        y_T &=& - c_t^2 s_t^2 \frac{v^2}{f^2}
	\label{eq:yT}
        \\
        y_{W_H} &=& -s^2 c^2 \frac{v^2}{f^2} 
                \nonumber \\
                &=& - \frac{M_W^2}{M_{Z_H}^2}
        \\
        y_{\Phi^+} &=& \frac{v^2}{f^2} 
        \left[ -\frac{1}{3} + \frac{1}{4} x^2 \right]
        \\
        y_{\Phi^{++}} &=& 0
                \label{eq:yPhi++}
        \\
        y_f &=& 1 + \frac{v^2}{f^2} \left[ - \frac{2}{3} + \frac{1}{2} x
                - \frac{1}{4} x^2 \right]
		\label{eq:yf}
        \\
        y_Z &=& 1 + \frac{v^2}{f^2} 
                \left[ -\frac{1}{6} - \frac{1}{4} (c^2-s^2)^2
                - \frac{5}{4} (c^{\prime 2}-s^{\prime 2})^2 + 
                \frac{1}{4} x^2 \right]
                \nonumber \\
                &=& 1 + \frac{v^2}{f^2} 
                \left[ -\frac{5}{3} + \frac{1}{4} x^2 \right]
                + \frac{M_W^2}{M_{Z_H}^2}
                + \frac{s_W^2}{c_W^2} \frac{M_W^2}{M_{A_H}^2}
        \\
        y^2_{M_Z} &=& 1 + \frac{v^2}{f^2} \left[ -\frac{1}{6} 
                - \frac{1}{4} (c^2-s^2)^2 
                - \frac{5}{4} (c^{\prime 2}-s^{\prime 2})^2
                + \frac{1}{2} x^2 \right]
                \nonumber \\
                &=& 1 + \frac{v^2}{f^2} \left[ -\frac{5}{3}
                + \frac{1}{2} x^2 \right]
                + \frac{M_W^2}{M_{Z_H}^2}
                + \frac{s_W^2}{c_W^2} \frac{M_W^2}{M_{A_H}^2}
        \\
        y^2_{M_W} &=& 1 + \frac{v^2}{f^2} \left[ -\frac{1}{6} 
                - \frac{1}{4} (c^2-s^2)^2 + \frac{1}{4} x^2 \right]
                \nonumber \\
                &=& 1 + \frac{v^2}{f^2} \left[ -\frac{5}{12}
                + \frac{1}{4} x^2 \right] + \frac{M_W^2}{M_{Z_H}^2}
        \\
        y^2_{c_W} &=& 1 + \frac{v^2}{f^2} \frac{s^2_W}{c^2_W-s^2_W}
                \left[ -\frac{1}{4} + \frac{1}{4} (c^2-s^2)^2 
                + \frac{5}{4} (c^{\prime 2}-s^{\prime 2})^2 
                - \frac{1}{4} x^2 \right]
                \nonumber \\
                &=& 1 + \frac{v^2}{f^2} \frac{s^2_W}{c^2_W-s^2_W}
                \left[ \frac{5}{4} - \frac{1}{4} x^2 \right]
                + \frac{s_W^2}{c_W^2 - s_W^2}
                \left[ -\frac{M_W^2}{M_{Z_H}^2}
                - \frac{s_W^2}{c_W^2} \frac{M_W^2}{M_{A_H}^2} \right].
                \label{eq:y2cW}
\end{eqnarray}
In Eq.~(\ref{eq:yPhi++}),
the $\Phi^{++}\Phi^{--}H$ coupling is zero at leading order in $v^2/f^2$
\cite{LHloop}, so the
corresponding $y_{\Phi^{++}}$ is suppressed by an extra factor of 
$v^2/f^2$ and we thus ignore it.
The $f\bar f H$ coupling in Eq.~(\ref{eq:yf}) was given previously
in Eq.~(B.10) of Ref.~\cite{Deandrea}; after correcting a 
typo~\cite{Deandreaprivate} we reproduce their result.

For the model in which only one U(1) group (hypercharge) is gauged, so that
there is no $A_H$ particle in the spectrum, the $y_i$ factors
in terms of the parameters $c_t$, $x$, $f$, and $c$ are obtained from
Eqs.~(\ref{eq:y2GF}--\ref{eq:y2cW}) by setting 
$c^{\prime} = s^{\prime} = 1/\sqrt{2}$.
The $y_i$ factors in terms of the parameters $c_t$, $x$, $f$, $M_{Z_H}$
are given as above except for
\begin{eqnarray}
        y_Z &=& 1 + \frac{v^2}{f^2} 
                \left[ -\frac{5}{12} + \frac{1}{4} x^2 \right]
                + \frac{M_W^2}{M_{Z_H}^2}
        \\
        y^2_{M_Z} &=& 1 + \frac{v^2}{f^2} 
                \left[ -\frac{5}{12} + \frac{1}{2} x^2 \right]
                + \frac{M_W^2}{M_{Z_H}^2}
        \\
        y^2_{c_W} &=& 1 + \frac{v^2}{f^2} \frac{s_W^2}{c_W^2 - s_W^2}
                \left[ - \frac{1}{4} x^2 \right]
                - \frac{s_W^2}{c_W^2 - s_W^2} \frac{M_W^2}{M_{Z_H}^2}.
\end{eqnarray}


\end{document}